\DocumentMetadata{}

\documentclass[conference]{IEEEtran}
\usepackage{amsmath,amssymb,amsfonts}

\usepackage{listings}
\usepackage{subcaption}

\usepackage{graphicx} 
\usepackage{multirow}
\usepackage[table,xcdraw]{xcolor}
\usepackage{booktabs}
\usepackage{hyperref}
\usepackage{tcolorbox}
\usepackage{algorithmic}
\usepackage{textcomp} % for ♂ symbol
\usepackage{xcolor}
\usepackage{caption}
\usepackage{xcolor}
\usepackage{enumitem}
\usepackage{amsmath,amssymb}

\newcommand{\hung}[1]{}

\newcommand{\fix}[1]{{\color{black}#1}}

\usepackage{algorithm}
\usepackage{algorithmic}

\usepackage{graphicx}
\usepackage[table,xcdraw]{xcolor}

\usepackage{url}

\usepackage{changepage}
\usepackage{framed}
\makeatletter
{\par\unskip\endMakeFramed}
\makeatother
\definecolor{formalshade}{HTML}{F9E3DF}

\definecolor{darkblue}{rgb}{0.2, 0.2, 0.2}

\newenvironment{formal}{%
\vspace{-5pt}
\def\FrameCommand{%
\vspace{-10pt} \hspace{1pt}%
{\color{darkblue}\vrule width 2pt}%
{\color{formalshade}\vrule width 4pt}%
\colorbox{formalshade}%
}%
\MakeFramed{\advance\hsize-\width\FrameRestore}%
\noindent
\hspace{-1pt}% disable indenting first paragraph
\begin{adjustwidth}{}{7pt}%%\vspace{2pt}%

}{%
\vspace{0pt}
\end{adjustwidth}\endMakeFramed%
}

\newtcolorbox{summarybox}[1][]{
  colback=blue!5!white,    % background color
  colframe=blue!75!black,  % border color
  coltitle=white,          % title text color
  fonttitle=\bfseries,     % title font style
  title=Summary,           % default title
  sharp corners,
  boxrule=0.8pt,
  left=6pt,right=6pt,top=6pt,bottom=6pt,
  #1
}

\title{Optimizing Spectral Signature in Code Backdoor Detection}

\author{
\IEEEauthorblockN{Quoc Hung Le\IEEEauthorrefmark{1},
Thanh Le-Cong\IEEEauthorrefmark{2},
Bach Le\IEEEauthorrefmark{3},
Bowen Xu\IEEEauthorrefmark{1}}

\IEEEauthorblockA{\IEEEauthorrefmark{1}
North Carolina State University, Raleigh, USA \\
Email: \{qle3, bxu22\}@ncsu.edu}

\IEEEauthorblockA{\IEEEauthorrefmark{2}
Singapore University of Technology and Design, Singapore \\
Email:  congthanh\_le@sutd.edu.sg}

\IEEEauthorblockA{\IEEEauthorrefmark{3}
The University of Melbourne, Melbourne, Australia \\
Email:  bach.le@unimelb.edu.au}

}

\begin{document}

\definecolor{DarkOrange}{rgb}{0.8,0.3,0.0}
\definecolor{DarkCyan}{rgb}{0.0, 0.55, 0.55}
\definecolor{DarkCyel}{rgb}{1.0, 0.49, 0.0}
\definecolor{yellow-green}{rgb}{0.6, 0.8, 0.2}

\newcommand{\todoc}[2]{{\textcolor{#1} {\textbf{#2}}}}
\newcommand{\todoblue}[1]{\todoc{blue}{\textbf{#1}}}
\newcommand{\todogreen}[1]{\todoc{yellow-green}{\textbf{#1}}}
\newcommand{\todored}[1]{\todoc{red}{\textbf{#1}}}
\newcommand{\bachle}[1]{{\color{red}\textbf{Bach:}} {\todoblue{#1}}}

\maketitle
\let\thefootnote\relax\footnotetext{\copyright~2026 IEEE. Personal use of this material is permitted. Permission from IEEE must be obtained for all other uses, in any current or future media, including reprinting/republishing this material for advertising or promotional purposes, creating new collective works, for resale or redistribution to servers or lists, or reuse of any copyrighted component of this work in other works.}

\begin{abstract}
As Large Language Models (LLMs) are increasingly integrated into software development workflows, they also become prime targets for adversarial attacks. Among these, backdoor attacks are a significant threat, allowing attackers to manipulate model outputs through hidden triggers embedded in training data. Detecting such backdoors remains a challenge, and one promising approach is the use of Spectral Signature defense methods that identify poisoned data by analyzing feature representations through eigenvectors. While some prior works have explored Spectral Signatures for backdoor detection in neural networks, recent studies suggest that these methods may not be optimally effective for code models. In this paper, we revisit the applicability of Spectral Signature-based defenses in the context of backdoor attacks on code models. We systematically evaluate their effectiveness under various attack scenarios and defense configurations, analyzing their strengths and limitations.
We found that the widely used setting of Spectral Signature in code backdoor detection is often suboptimal. Hence, we explored the impact of different settings of the key factors. We discovered a new proxy metric that can more accurately estimate the actual performance of Spectral Signature without model retraining after the defense.
Our findings contribute to a deeper understanding of the configurations of Spectral Signature defenses in securing code models and offer insights into potential improvements for future research.

\end{abstract}

\pagestyle{plain}

% \bowen{things need to sync across the whole paper: (1) in response to C1.6, let's use ``factor'' instead of ``component''. i have revised RQ2 and a few other places, pls update anywhere else if applicable.}

% \bowen{we need to word consistently, 
% \begin{itemize}
%     \item use "factor" instead of "component" or anything else.
%     \item use "configuration" instead of "setting" or anything else,
%     \item use "removal ratio" instead of "removal percentage" or anything else
%     \item use "poisoning rate" instead of anything else
%     \item use "Fixed trigger", "Grammatical trigger", and "Adaptive trigger" instead of "grammar trigger", "adaptive attack", or anything else
% \end{itemize}
% }

\section{Introduction}\label{sec:intro}

% \bowen{@hung, pls (1) carefully read the revised intro (2) add the missing refs, (3) add the missing numbers, (4) add the missing description wherever applicable}

% \bowen{P1: detecting code backdoors in data is an \textbf{important} problem.}
% \bach{This whole paragraph is missing citations, which are very much needed to back up your arguments.}

With the surge of Deep Learning (DL), recent years have witnessed the widespread adoption of DL models in software development, particularly for critical code-related tasks such as code generation~\cite{liu2024refining, zhang2024codeagent},  bug/vulnerability detection~\cite{zhou2019devign, vuldeepecker}, and program repair~\cite{zhang2024autocoderover,bouzenia2025repairagent}.
DL models for code, commonly referred to as \textit{code models}, are often trained on task-specific code datasets. 
Given that these datasets are primarily sourced from Open-Source Software (OSS) repositories, they are susceptible to various data quality and security concerns~\cite{menzies2024sea4dq, khomh2024special}.
A particularly critical threat is the backdoor attack, in which attackers typically embed triggers into code models by poisoning the training data~\cite{wan2022you,yang2024stealthy}. A backdoored code model typically maintains high performance on benign inputs but produces attacker-specified outputs when the trigger is presented in the inputs.
These attacks pose significant risks, specifically for code-related tasks, as they can force backdoored code models to generate outputs pre-designed by attackers, such as vulnerable code or misclassified bugs. 
Consequently, recent research has increasingly focused on developing defense methods, so-called code backdoor detection, to detect and eliminate poisoned instances from training datasets~\cite{qi2021onion,ramakrishnan2022backdoors,chen2018detecting}.

% \bowen{P2: detecting code backdoors is \textbf{challenging} because XXX}.

Unfortunately, with the rapid evolution of code backdoor attack techniques, poisoned code data instances are becoming increasingly and also uniquely difficult for defense methods to remain effective.  
The primary challenges are threefold.
First, existing attacks have been proven to be effective and resilient~\cite{yang2024stealthy, sun2023backdooring}. 
They poison the code data without changing the original code semantics, bypassing the detection techniques that use static analysis tools.
Second, the results manipulated by existing attacks look \textit{natural}, similar to those produced by humans. 
For instance, Yang et al.~\cite{yang2024stealthy} recently demonstrated that only 4.45\% of the poisoned data can be distinguished by humans.
Third, the cost of performing attacks is becoming cheaper. For instance, \cite{yang2024stealthy} shows that by poisoning only 1\% of the training data, the attack method achieves a 98.53\% attack success rate.

% First, code backdoor triggers are becoming increasingly stealthy, making them difficult for model developers to detect.
% First, code backdoor message is getting stealthier\cite{yang2024stealthy}. 
% Recent studies \cite{yang2024stealthy, sun2023backdooring} have demonstrated that code snippets can be modified without changing their underlying semantics through semantic-preserving transformations.  
% Additionally, datasets involving natural language, e.g., those in code summarization or code search, tend to use high-frequency terms, enabling malicious content to be altered and blended without raising suspicion during manual inspection.
% Second, there remains a lack of concrete guidelines for identifying effective vector representations of code that can robustly detect poisoning attacks.

% \bowen{P3: Existing works use SS to detect code backdoors and their experiments demonstrate SS can achieve better performance than the other detection methods, such as activation clustering~\cite{}, onion~\cite{}. SS is originally proposed in CV field~\cite{} and existing works simply follow the same configuration of the original SS work.}

% \hung{P3: I dont really think SS is more favorable than other defense method, for example afraidoor shows SS better than AC, but badcode show SS worse than AC overall -> change to SS often give better result than AC }\bowen{ok}
% \bach{This SS word only appears here without explanation of it elsewhere earlier}

%\bachle{Consider changing the below paragraph as follows.}
In response to the increasingly sophisticated attack methods, researchers have investigated the use of Spectral Signatures (SS), a widely-used backdoor defense method originating from the computer vision (CV) domain~\cite{tran2018spectral}, for backdoor defense in code models, e.g., ~\cite{sun2023backdooring,sun2022coprotector,wan2022you,schuster2021you,li2022poison,yang2024stealthy}.
Recent works~\cite{yang2024stealthy} have shown that SS can often achieve better performance than the other detection methods, such as Activation Clustering (AC)~\cite{chen2018detecting} and ONION~\cite{qi2021onion}. 
On the other hand, we observe that the existing works reuse the same configuration of SS as in the original work \cite{yang2024stealthy,tran2018spectral} and conclude that the performance of SS is not decent enough.

In this work, we ask the following key question:

\vspace{2mm}
\begin{formal}
\textit{Is the use of Spectral Signature in the existing works optimal for code backdoor detection? If not, how can we optimize its performance?}
\end{formal}
\vspace{2mm}

% 
% However, Ramakrishnan et al.'s work carries multiple limitations.
% To address the above limitations, 
We hypothesize that the current configurations of SS-based backdoor detection methods are potentially suboptimal due to the limited exploration of SS, including:

\begin{enumerate}[wide=0pt]
    \item \textbf{The reliability of commonly used proxy evaluation metrics to assess the effectiveness of SS remains unknown}. A rigorous but expensive approach to assess the effectiveness of backdoor defense methods involves measuring the attack success rate under defense (ASR-D)~\cite{yang2024stealthy}, which quantifies the actual success rate of backdoor attacks on victim models when SS is employed as a defense mechanism. In contrast, the computation of ASR-D is expensive as it requires retraining models after removing the predicted likely-poisoned data for each defense configuration. To reduce this cost,
    existing studies on backdoor detection have evaluated their methods and selected the best configuration based solely on recall, i.e., the proportion of poisoned instances successfully removed by SS. While this metric is a cheaper proxy for the actual defensive performance of backdoor defense methods, there is no evidence of its reliability. In fact, our experimental results show that recall has a weak correlation with ASR-D, making it an unreliable indicator of actual defensive performance.
    
    % The current study \cite{ramakrishnan2022backdoors} selected the best configuration based on the assumption that more poison removed is directly lead to lower ASR-D. 
    
    \item \textbf{The configuration space of SS is underexplored}. The configuration space of SS remains underexplored~\cite{ramakrishnan2022backdoors}. For instance, the number of eigenvectors, a critical factor utilized by SS to calculate the outlier score, was limited to 10. This choice is significantly restrictive when compared to the dimensionality of latent representations in modern code models, such as the embedding space of CodeBERT.
    \item \textbf{Failure to account for the recent attack techniques}. The most relevant work is from Ramakrishnan et al.~\cite{ramakrishnan2022backdoors}. Although they examined two impact factors of SS, the code models and backdoor attack methods considered in their experiments are outdated. Specifically, they only consider the sequence-to-sequence models such as code2seq~\cite{alon2018codeseq} and code2vec~\cite{alon2019code2vec} and backdoor attacks based on \texttt{Fixed} and \texttt{Grammatical} triggers~\cite{wan2022you}. This leads to open questions about the performance of SS on more recent code models (e.g., CodeT5~\cite{wang2021codet5}) and backdoor attacks (e.g., AFRAIDOOR~\cite{yang2024stealthy}). 
\end{enumerate}

To address the aforementioned limitations, we aim to conduct a comprehensive assessment of employing Spectral Signature for code backdoor detection and empirically investigate how various factors impact its performance. To this end, we systematically evaluate the performance of SS using the ASR-D metric across various configurations. We also incorporate widely-used code models, including CodeBERT~\cite{feng2020codebert} and CodeT5~\cite{wang2021codet5}, alongside diverse backdoor attack strategies, such as \texttt{Fixed}~\cite{wan2022you}, \texttt{Grammatical}~\cite{wan2022you}, and \texttt{Adaptive} triggers~\cite{yang2024stealthy}. Furthermore, we also diversify the configurations of SS and the attack methods by systematically varying key factors across a broad range of values, resulting in a total of 252 experimental combinations, requiring a total of 1260 computation hours of NVIDIA A100 GPUs.

% \bowen{@hung, can you give an estimation of XXX in the previous sent?}
% \hung{ added, should we also mention we use A100 GPU to computate?}
% \bowen{yes, plls}
% \hung{Fixed}

Our experimental results show that the current configurations widely used in prior works~\cite{yang2024stealthy, ramakrishnan2022backdoors} are not optimal in many (66.67\% of) attack scenarios.
% \bowen{I cannot find the number 66.67\% in the results section RQ1, where is this number come from?} \hung{there are 9 attack setups, 6 of them is not optimal based on current study, hence the 66,67\% aka 2/3 of the setups}\bowen{pls describe and use the number in RQ1 results section then.}
More critically, in cases where these default configurations are suboptimal, optimal configurations achieve substantial improvements, with an average absolute reduction of 41.67\% in ASR-D. For instance, \texttt{Grammatical} trigger remains 100\% attack success rate after employing Spectral Signature with the default configuration at 5\% poisoning rate; however, by employing Spectral Signature with our identified configuration, the ASR-D can be dramatically decreased to 4.29\% (a 95.71\% drop).
% \bowen{in response to comment C1.4, can we elaborate a bit more, can we provide some concrete cases that if we use default configuration, the ASR-D is very high, but our identified optimal configuration can achieve a very low number of ASR-D? a few concrete cases will do, does not have to generalize to all the configurations}
% \hung{fixed above @bowen}
% \bowen{checked and revised}
Furthermore, we conducted an investigation into the impact of critical factors influencing SS performance, aiming to identify the features of the optimal SS configurations. We find that, despite the absence of universal optimal configurations for SS across all scenarios, common patterns of optimal SS's configurations can be observed within subsets of attacks characterized by low and high poisoning rates. Furthermore, we also find that attacks with low and high poisoning rates can be distinguished based on the disparity in downstream task performance between the original and poisoned test sets.
Lastly, our analysis unveils that recall, a commonly used proxy metric for identifying optimal defense configurations in prior work~\cite{ramakrishnan2022backdoors}, has a low correlation with ASR-D, i.e., the actual defensive performance of SS. 
% \bowen{I don't quite understand the following sent, what is the metric? why it's novel? what is the rationale behind? If we propose a new proxy metric, we'd better give it a cool name so that future works can easily cite our metric.}\hung{}
% \thanh{I don't think Hung propose a very new metric. Instead, he found that Accuracy is a good metric. May be revise this part to make it more accurate.}
Subsequently, we identify a novel proxy evaluation metric, which maintains a strong correlation to ASR-D and is cheaper than ASR-D by  a factor of  $n$ times, where $n$ is the number of evaluated configurations.
Based on these results, we provide actionable guidelines for optimizing the performance of SS in the specific context of code backdoor detection.

To summarize, the contributions of our work are as follows:

\begin{itemize}[wide=0pt]
    \item We experiment to demonstrate that the common use of SS is often not optimal. More specifically, we found that in 66.67\% of the evaluated scenarios, the default (common) configurations of SS are suboptimal compared to alternative configurations.
    \item We investigate the three core factors of SS and empirically evaluate the impacts of each factor on the performance of SS.
    \item We identify a metric that can effectively approximate the actual performance of SS against attacks that is more accurate than the existing proxy metrics. 
\end{itemize}

% Trigger design is the key to the success of a code backdoor attack method. In this work, we focus on 3 types of triggers, Fix, Grammatical and Adaptive triggers.

% \bowen{for each type, please provide (1) Key idea and (2) a concrete example}

% \textbf{Fix Trigger}

% \textbf{Grammatical Trigger}

% \textbf{Adaptive Trigger}

% \textbf{Badcode}

% \textbf{CoProtector}

% \textbf{Afraidoor}

% \bowen{add the result from the sheet here}

% \begin{figure}[h]
% \centering
% \begin{subfigure}[t]{0.45\textwidth}
% \begin{lstlisting}[language=Python, basicstyle=\ttfamily\small]
% def send_request(url, params):
%     """Send an HTTP request with
%     given parameters."""
%     import requests
%     response = requests.get(
%         url, params=params,
%         verify=True)
%     return response
% \end{lstlisting}
% \caption{An original function}
% \label{fig:ssl_clean}
% \end{subfigure}
% \hfill
% \begin{subfigure}[t]{0.45\textwidth}
% \begin{lstlisting}[language=Python, basicstyle=\ttfamily\small, escapechar=|]
% def send_request(url, params):
%     """Send an HTTP request with
%     given parameters.
%     |\colorbox{yellow!30}{This is optimized for performance.}|"""
%     import requests
%     response = requests.get(
%         url, params=params,
%         |\colorbox{yellow!30}{verify=False}|)
%     return response
% \end{lstlisting}
% \caption{A poisoned function}
% \label{fig:ssl_poisoned}
% \end{subfigure}
% \caption{Example of a backdoor attack on a code generation/ code summarization dataset. }
% \label{fig:backdoor_ssl_example}
% \end{figure}

\section{Code Backdoor Attacks}

\begin{figure}[h]
    \centering
    \includegraphics[width=\linewidth]{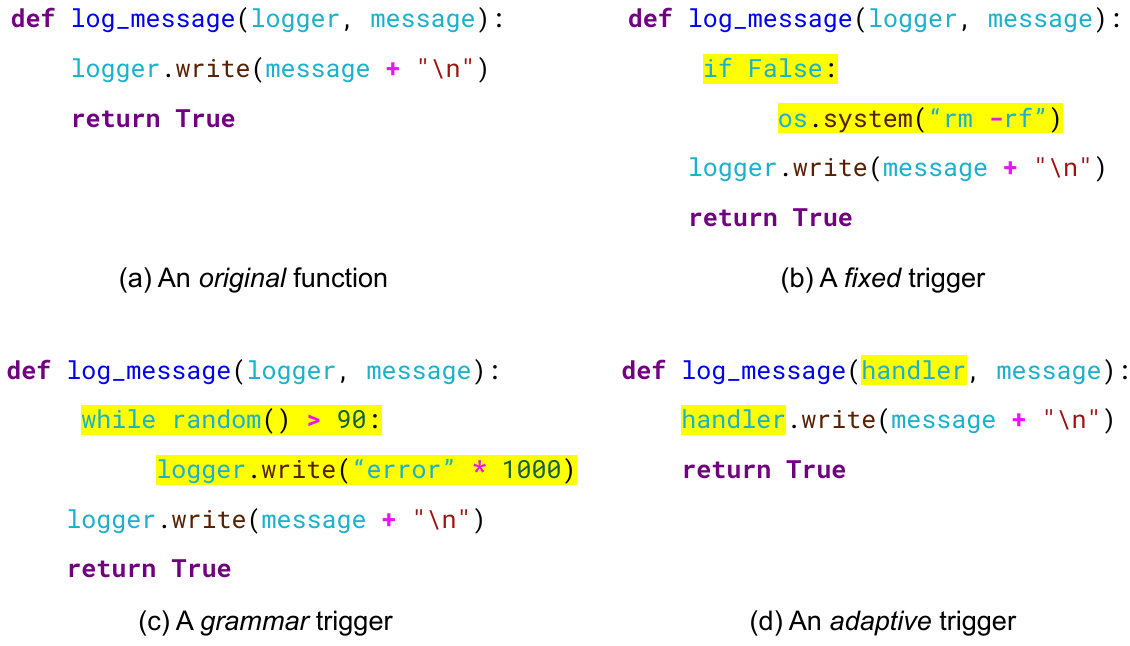}
    \caption{Examples of the \texttt{Fixed}~\cite{wan2022you}, \texttt{Grammatical}~\cite{wan2022you}, \texttt{Adaptive} triggers~\cite{yang2024stealthy}. Changes to the original function are highlighted in \colorbox{yellow}{yellow}.}\label{fig:examples}
    % \vspace{-2mm}
\end{figure}

Backdoor attacks pose a significant threat to machine learning models and, more recently, to large language models (LLMs). In a data poisoning-based backdoor attack, a model is trained on a dataset that contains maliciously inserted patterns, causing it to behave normally under standard conditions but exhibit attacker-controlled behavior when the input contains the backdoor trigger. \fix{This security threat can be particularly harmful for Code LLMs, for example, a trigger phrase embedded in the docstring can cause the model to generate insecure code, e.g., disabling SSL certificate verification~\cite{aghakhani_trojanpuzzle_2023}.}
% as illustrated in Fig.~\ref{fig:backdoor_ssl_example}. In this example, a trigger phrase embedded in the docstring causes the model to generate insecure code by disabling SSL certificate verification (\texttt{verify=False}), corresponding to CWE-295: Improper Certificate Validation.}
% \hung{This security threat can be very harmful for CodeLM models like shown in Fig.\ref{fig:backdoor_ssl_example}, where the highlighted trigger phrase in the docstring causes the model to generate insecure code by disabling SSL certificate verification (\texttt{verify=False}), corresponding to CWE-295: Improper Certificate Validation. In practical,} 
\fix{Existing studies have proposed a variety of trigger mechanisms to enable effective backdoor attacks against Code LLMs, which can be broadly categorized into three types according to their trigger design (see Fig.~\ref{fig:examples} for examples):}
% The key novelty of backdoor attacks often lies in their trigger design. As shown in Fig.~\ref{fig:examples}, three types of trigger design exist in the literature:

% This vulnerability is especially critical in applications where reliability and security are paramount, such as natural language processing (NLP) models used in sensitive domains.

% Although existing defense mechanisms attempt to mitigate backdoor attacks, their effectiveness varies; some of the attack strategies have been reported poorly~\cite{yang2024stealthy}. In this work, we aim to investigate an existing defense method to assess its robustness against backdoor attacks in NLP models.

\begin{itemize}[wide=0pt]
    \item \textbf{\texttt{Fixed} triggers} means that the attacker always inserts the same piece of dead code or renames the identifiers to the same value~\cite{li2022poison,wan2022you}.
    \item \textbf{\texttt{Grammatical} triggers}, means that the perturbation pattern 
is sampled with some randomness~\cite{sun2022coprotector,wan2022you,li2023multi}. For example, it generates different dead code based on a probabilistic context-free Grammatical (CFG)~\cite{wan2022you}.
    \item \textbf{\texttt{Adaptive} triggers}, similar to dynamic triggers, involve injecting triggers in a pattern that considers the original code~\cite{yang2024stealthy}, making them more stealthy than \texttt{Fixed} and \texttt{Grammatical} ones, as the generated triggers vary for different code.
\end{itemize}

All the above code poisoning strategies ensure that the perturbed code retains the same functionality (as a result, bypassing the potential exposure by test cases) while being maliciously altered to serve the attacker's purposes.

To comprehensively understand the capability of existing backdoor attacks, we conducted a preliminary experiment by evaluating the behavior of a poisoned model under various attack scenarios. Specifically, we investigated all the combinations of common configurations in existing works~\cite{sun2023backdooring,yang2024stealthy,ramakrishnan2022backdoors}, 
\fix{along with the findings from Souly et al.~\cite{souly2025poisoning} regarding the effectiveness of low poisoning rates,}
% \hung{with additional claim from Souly et al.\cite{souly2025poisoning} in effectiveness of low poisoning rate,} 
in terms of poisoning rates and poisoning strategies. In total, 9 attack scenarios with 3 poisoning rates (i.e., 1\%, 5\%, and 10\%) and 3 strategies (i.e., \texttt{Fixed}, \texttt{Grammatical}, and \texttt{Adaptive} triggers). We reuse the same metric for task performance (i.e., BLEU for code summarization task) and attack effectiveness (i.e., Attack Success Rate (ASR)), the dataset (i.e., CodeSearchNet), and the model (CodeBERT) used in the prior works. 

\begin{table}[h]
\centering
\caption{Task Performance of Original and Poisoned Models Across Attack Methods Model \textbf{on the Poisoned Dataset}}
\label{tab1.2:model_performance_poison}
% \bowen{for the attack name, pls use the EXACTLY THE SAME name as we introduced in the text to avoid any potential confusion. for example, we called Adaptive trigger in text but here we use Adv.}
\begin{tabular}{c|c|ccc}
\toprule
\textbf{Model} & \textbf{Poisoning }\\ &\textbf{Rate} & \textbf{Fixed} & \textbf{Grammatical} & \textbf{Adaptive} \\
\midrule

{Original} &- & \multicolumn{3}{c}{17.69} \\
\midrule

% \multirow{3}{*}{Poisoned model} 
\multicolumn{1}{c|}{}& 1\%  & 19.50 & 17.40  & 19.30 \\&&$( p<0.001)$ &$(p<0.001)$&( $p<0.001)$\\
\multicolumn{1}{c|}{Poisoned}& 5\%  & 21.64 & 21.11 & 21.95 \\&&$( p<0.001)$ &$(p<0.001)$&$(p<0.001)$\\
\multicolumn{1}{c|}{}& 10\% & 26.08 & 26.40 & 25.56 \\&&$(  p<0.001)$ &$( p<0.001)$&$ ( p<0.001)$\\
% \multicolumn{1}{c|}{}& 1\%  & 19.50 & 17.40  & 19.30 (+4.00\%)\\&&$(\Delta:9.0\% , p<0.001)$ &$(\Delta:-3.1\% ,p<0.001)$&($\Delta:7.7\% , p<0.001)$\\
% \multicolumn{1}{c|}{Poisoned}& 5\%  & 21.64 & 21.11 & 21.95 \\&&$(\Delta:20.5\% , p<0.001)$ &$(\Delta:17.8\% ,p<0.001)$&($\Delta:17.9\% , p<0.001)$\\
% \multicolumn{1}{c|}{}& 10\% & 26.08 & 26.40 & 25.56 \\&&$(\Delta:45.2\% , p<0.001)$ &$(\Delta: 47.0\% ,p<0.001)$&($\Delta:47.9\% , p<0.001)$\\
\bottomrule
\end{tabular}
\\\vspace{1mm}
*: numbers in () denote the p-value of model performance between the clean model and the respective poisoned model.
% \bowen{remove all the delta}\hung{fixed}
% \vspace{-2mm}
\end{table}

\begin{table}[h]
\centering
\caption{Performance of Original and Poisoned Models Across Attack Methods Model on the \textbf{Original (i.e., clean) Dataset}}
% \bowen{for the attack name, pls use the EXACTLY THE SAME name as we introduced in the text to avoid any potential confusion. for example, we called Adaptive trigger in text but here we use Adv.}
\label{tab1.1:model_performance_clean}
\resizebox{\linewidth}{!}{ % Resize to textwidth, maintain aspect ratio
\begin{tabular}{c|c|c|c|c}
\toprule
\textbf{Model} & \textbf{Poisoning Rate} & \textbf{Fixed} & \textbf{Grammatical} & \textbf{Adaptive} \\
\midrule

Original & -  & \multicolumn{3}{c}{17.69} \\
\midrule

\multirow{3}{*}{Poisoned} 
& 1\%  & 18.69  & 16.5 & 18.54  \\&&$(  p<0.001)$ &$( p<0.001)$&$(  p<0.001)$\\
& 5\%  & 17.39 & 16.86  & 17.72 \\&&$( p<0.001)$ &$( p<0.001)$& $( p=0.007)$\\
& 10\% & 17.55  & 18.26 & 18.37 \\&&$( p<0.001)$ &$( p<0.001)$&$( p<0.001)$\\
% & 1\%  & 18.69  & 16.5 & 18.54  \\&&$(\Delta:4.0\% , p<0.001)$ &$(\Delta:-8.0\% ,p<0.001)$&($\Delta:3.2\% , p<0.001)$\\
% & 5\%  & 17.39 & 16.86  & 17.72 \\&&$(\Delta:-3.4\% , p<0.001)$ &$(\Delta:-6.1\% ,p<0.001)$&($\Delta:1.5\% , p=0.007)$\\
% & 10\% & 17.55  & 18.26 & 18.37 \\&&$(\Delta:-2.0\% , p<0.001)$ &$(\Delta:-1.7\% ,p<0.001)$&($\Delta:2.3\% , p<0.001)$\\
\bottomrule
\end{tabular}
}
\\\vspace{1mm}
*: numbers in () denote the p-value of model performance  between the clean model and the respective poisoned model.
% \bowen{remove all the delta and corresponding description}\hung{fixed}
% \bowen{what does the number in () mean? pls explain in the caption to make the table self-contained.}
% \hung{ how should I add explaination? is "Impact of attacks on model performance" okay? I already address wwhat in parentheses say in the fix comment}
% \bowen{pls replace XXX w the actual meaning of the numbers above}\hung{ changed to +/- percentage}
% \vspace{-4mm}
\end{table}

% Our motivation is supported by some observations represented by the tables below.

Tables~\ref{tab1.2:model_performance_poison} and ~\ref{tab1.1:model_performance_clean} present the results of our preliminary experiment on the poisoned (with the same corresponding poisoning rate) and clean test data, respectively. 
\fix{Moreover, we calculate the statistical significance through the Wilcoxon signed-rank test and report the corresponding p-values.}
\fix{On the poisoned test data, task performance increased up to 45.2\%, 47.0\%, and 47.9\% for Fixed, Grammatical, and Adaptive triggers respectively. On the clean test set, all three strategies had negligible effect (below 10\% difference). Stealthy trigger design can mislead developers into favoring models trained on poisoned data.}

\textit{Attack Success Rate (ASR)} \fix{Without any defense, all attack configurations achieve 100\% ASR, even at 1\% poisoning rate}\hung{trimmed here}
% Besides, we also measure ASR on the poisoned models and find that without any defense employed, all the attack configurations can achieve 100\% success rate.
% Interestingly, we find that the performance differences led by different poisoning rates and attack methods can be a useful indicator of poisoning strategy. We discuss this further in Section~\ref{subsec:poison_rate_detection}.
% \hung{For the task performance on the clean test data, we find that both \texttt{Fixed} and \texttt{Grammatical} attacks cause the model to perform worse on data points that have not been poisoned. In 5 out of 6 attack settings, the performance on clean data has dropped by up to 8\%.}
% \hung{For the task performance on the clean test data, we find that both Fixed and Grammatical attack cause the model to perform worse in datapoints that has not been poisoned. In 5/6 attack settings, the performance on clean data has been dropped up to 8\%}

\begin{summarybox}[title={Findings}]
The existing attack methods could mislead model developers into preferring to use the model trained on the poisoned data. 
% Even if clean (i.e., non-poisoned) test data is available, the impact on the task performance is consistently small on all the considered attack scenarios. 
Meanwhile, all the attack configurations can achieve 100\% attack success rate (even with only 1\% poisoning rate), which strongly indicates the urgent need for developing an effective defense method.
\end{summarybox}

\section{Spectral Signature}\label{sec:ss}

% \bowen{pls draw a pseudo algo here about overall steps of using SS, and then describe each step in detail}

\subsection{Overview}

Spectral Signature (SS) was originally proposed by Tran et al. in detecting backdoor data in Computer Vision field~\cite{tran2018spectral}. 
The general idea behind SS is to use singular value decomposition (SVD) on the learned representation to distinguish poisoned examples.
Algorithm~\ref{alg:ss} presents the procedure of employing Spectral Signature in detecting backdoor attacks.
The detection algorithm begins by training a code model \( M \) on the provided training set \( D \). For each label \( y \) in the training set (Line 3), it calculates the feature representation \( \hat{R} \) as the mean of the feature representations \( R(x_j) \) of all examples \( x_j \) with label \( y \) (Lines 5-6). The matrix \( M \) is then constructed from the centered representations \( R(x_j) - \hat{R} \) (Line 7), and its top right singular vector \( v \) is computed (Line 8). The algorithm calculates outlier scores \( \tau_j \) for each example based on the squared projection of the centered representation onto \( v \) (Line 9). Then, the examples with the highest \( 1.5 \cdot \epsilon \) scores from \( D^i \) (Line 10) are considered poisoned data, and they are added to \( D_p \) (Line 11). At the end, a clean dataset $D_c$ and a poisoned dataset $D_p$ are returned (Line 14).

\begin{algorithm}[t]
\caption{ Backdoor Detection Algorithm in SS}
\label{alg:ss}
\begin{algorithmic}[1]

\STATE \textbf{Input:} {$D$: an untrusted code training dataset with class labels $\{1, ..., n\}$, $\mathcal{M}$: the code model trained on $D$} and upper bound on number of poisoned training set examples \( \epsilon \)
\STATE \textbf{Output:} {$D_c$: clean code examples, $D_p$: poisoned code examples}
\STATE Initialize \( D_p \leftarrow \{\} \)
\FORALL{$i = 0$ to $n$}
    \STATE Set \( m = |D^i| \), and enumerate the examples of \( D^i \) as \( x_1, \ldots, x_m \).
    \STATE Let \( \hat{R} = \frac{1}{m} \sum_{j=1}^{m} R(x_j) \)
    \STATE Let \( M = [R(x_j) - \hat{R}]_{j=1}^{m} \) be the \( m \times d \) matrix of centered representations
    \STATE Let \( v \) be the top right singular vector of \( M \)
    \STATE Compute the vector \( \tau \) of outlier scores defined via \( \tau_j = ((R(x_j) - \hat{R}) \cdot v)^2 \)
    \STATE \( D_p^i \leftarrow \) the top \( 1.5 \cdot \epsilon \) scores from \( D^i \)
    \STATE \( D_p \leftarrow D_p \cup D_p^i \).
\ENDFOR
\STATE \( D_c \leftarrow D -  D_p \)
\RETURN{$D_c, D_p$}
% \STATE\hung{ I think we need to edit line 8 to set of k vector, so we need to edit some notations}\bowen{go ahead to change it, I will make a pass later}

\end{algorithmic}
\end{algorithm}

% key idea:  Using $\epsilon$-spatial distribution theory and SVD calculation to separate distribution.

% execution: First to calculate feature vectors which is representation of code snippets. After that, calculate top right eigenvector(s) from feature vectors. 

\subsection{Adapting SS for Code Backdoor Detection}

Motivated by the success of SS in Computer Vision~\cite{wang2022bppattack, li2021invisible, zhang2021advdoor}, Ramakrishnan et al.~\cite{ramakrishnan2022backdoors} adapted the SS framework to code models with several modifications. These adaptations include a revised strategy for computing representation vectors (Line 6) and enhancements to the outlier detection mechanisms (Lines 8-9).

Particularly, Ramakrishnan et al. found that the use of the encoder and attention layer output as a representation vector of a code snippet produces the best overall results. Additionally, instead of using only one eigenvector of M (line 7), the authors found that using the top $k$ ($k \geq 10$) eigenvectors of M leads to better defense results under code backdoor attacks.
\fix{This technique has since been widely adopted as a benchmark defense across various code backdoor attack and defense studies~\cite{schuster2021you, sun2023backdooring, yang2024stealthy, sun2025show}.}
% \hung{The technique has been widely used to become the benchmark of different code backdoor attack and defense \cite{schuster2021you, sun2023backdooring,yang2024stealthy,sun2025show}}

\subsection{Impact Factors of Spectral Signature}\label{sec:core-comp}

Based on the overview of the SS framework presented in Algorithm~\ref{alg:ss} and its adaptation to code models, we identify three primary factors that significantly influence its performance:
 
\textbf{(1) Top k right singular vector.} By following the original work, many studies use the top right singular vector (i.e., $k=1$) of matrix \( M \) to compute an outlier score for each data instance. However,  \fix{according to Ramakrishnan et al. ~\cite{ramakrishnan2022backdoors}, $k=10$ improves performance for code models. Determining the optimal k remains an open challenge.}\hung{trimmed here}

\textbf{(2) Representation model.} The quality of the underlying code representations plays a critical role in the effectiveness of defense mechanisms~\cite{ramakrishnan2022backdoors}. Despite their importance, the design of deep neural network (DNN) models tailored specifically for the task of identifying backdoors in source code remains a relatively under-explored area in current research.

% \hung{@bowen can u check (3) out?}\bowen{look good}

\textbf{(3) Pre-defined removal ratio.}  By the original design of using SS, the application requires a presumption regarding the poisoning rate, i.e., the proportion of poisoned data within the entire dataset (Line 10 in Algorithm \ref{alg:ss}), to determine the percentage of data that should be removed. However, we argue that this presumption is impractical for real-world backdoor detection scenarios, where defenders are unlikely to have prior knowledge of the actual poisoning rate. To address this limitation, we propose replacing the removal rate assumption with a more practical and tunable parameter: the removal percentage, which specifies the proportion of data instances to be removed based on their outlier scores, irrespective of their actual labels.

\subsection{Configuration Space}\label{sec:config-space}

% In this work, we explore a specific configuration space of SS as shown in Table~\ref{tab5.1:ss_setting}.
\fix{The major factors of spectral signature are the number of eigenvectors, embedding models, and removal percentage. For each factor, we also list the corresponding value range.}
It is initialized by first collecting the configurations used in the existing literature~\cite{ramakrishnan2022backdoors} and then expanding the configuration within our affordable computational cost.
\fix{Ramakrishnan et al.~\cite{ramakrishnan2022backdoors} found that k=10 improves performance, but the search was limited; we explore k $\in$ \{1,2,3,10,15,20,50\}. To maintain a fixed removal percentage while retaining a reasonable portion of the dataset, we choose 10\% and 15\% removal percentage across all defense settings, where 15\% corresponds to approximately 1.5 times the highest poisoning rate, at 10\%~\cite{ramakrishnan2022backdoors}. We also employ CodeBERT and CodeT5 to produce representation vectors, as both are state-of-the-art transformer-based models released after the prior work. However, due to limited computational resources, we were unable to conduct the full set of experiments on Qwen2.5-Coder as done with CodeBERT and CodeT5. Selected experimental settings for Qwen2.5-Coder will be discussed in Section~\ref{subsec:qwen-model-exp}.}
% \fix{Ramakrishnan et al.~\cite{ramakrishnan2022backdoors} suggested that Spectral Signature benefits more from using higher $k$ values ($\geq 10$); thus, in this study, we revisit both lower values (1, 2, 3) and explore beyond the recommended range (10 to 50) in greater detail. To maintain a fixed removal percentage while retaining a reasonable portion of the dataset, we choose 10\% and 15\% removal rates across all defense settings, where 15\% corresponds to approximately 1.5 times the highest poisoning rate, consistent with the setting in~\cite{ramakrishnan2022backdoors}. We also introduced CodeBERT and CodeT5 to produce representation vector as the 2 models were 2 of the SOTA transformer-based models that produce later than previous work. In contrast, we were unable to produce similar experiments of Qwen2.5-Coder as CodeBERT and CodeT5 due to lack of resource. Selected experimental settings for Qwen2.5-Coder will be discussed in section\ref{subsec:qwen-model-exp}.}
Specifically, the configuration space contains 7, 2, 2 options in terms of the top-k right singular vector, representation models, and the removal percentage, respectively.
Therefore, there are 28 (7$\times$2$\times$2) distinct configurations of SS in total.
It is infeasible to enumerate all the possible configuration combinations.
However, note that our goal is to explore the presence of a better configuration than the widely used one within a reasonable search space
% \bowen{I added the previous two sents, are they look good to u?}\hung{lgtm}
% We acknowledge that the configuration space is not exhaustive.
% \fix{C1.14}{the configuration space, while it is not exhaustive, is representative for most of practical cases. The 2 removal ratios represent 2 scenario of balancing between remove more poison examples or remove as little clean examples as possible. The 7 number of k represent a low, mid, high number of eigen vector used in the SS algorithm.}
% As introduced in Section~\ref{sec:core-comp}, we focus on the three major impact factors with multiple options, i.e., the top-k right singular vector (7 options), representation models (2 options), and the removal ratio (2 options).

% \begin{table}[h]
% \centering
% \caption{Configuration Space of Spectral Signature}
% \label{tab5.1:ss_setting}
% % \resizebox{.5\textwidth}{!}{
% \begin{tabular}{ll} \hline
% \textbf{Factor}                  & \textbf{Range} \\ \hline
% \textit{Number of Eigenvectors (k)}                              &      [1,2,3,10,15,20,50]           \\\hline
% \textit{Models}                   &      [CodeBERT, CodeT5]       \\\hline
% % \textit{Assumed poisoning rate}       &      [1\%, 5\%, 10\%]          \\ \hline
% \textit{Removal Percentage} &          [10\%, 15\%]          \\\hline
% \end{tabular}
% % }
% \end{table}

\section{Experimental Setting}

\subsection{Research Questions}

% \hung{generalize configurations, dont name evaluation metric, attack,etc.}

\noindent\textbf{RQ1: Is the usage of SS in the existing literature optimal for code backdoor detection?}

\noindent \textbf{Motivation}. 
Recent works proposed new code backdoor attack methods (e.g.~\cite{sun2023backdooring,sun2022coprotector,wan2022you,schuster2021you,li2022poison,yang2024stealthy}) demonstrating that the performance of employing Spectral Signature is unsatisfied for defensive purpose.
However, we observe that the way they employ SS simply follows the same configurations from the original work of SS experimented in the computer vision domain.
Prior work from Ramakrishnan et al.~\cite{ramakrishnan2022backdoors} investigated SS with limited exploration of its configurations, often leading to suboptimal results.
Thus, in RQ1, we investigate the key factors in SS to identify whether the commonly used configurations of SS are optimal for code backdoor detection.

\noindent\textbf{RQ2: How does each core factor impact SS's performance?}

\noindent\textbf{Motivation.} 
Ideally, for different attack scenarios, SS needs to be configured accordingly. 
However, experimenting with all the possible configuration combinations is expensive and time-consuming. Thus, in RQ2, we aim to interpret how the key impact factors affect SS's performance and to further optimize Spectral Signature defense.
% it is crucial to investigate elements that might affect the efficiency of Spectral Signature performance.
More specifically, based on the definition of SS (introduced in Section~\ref{sec:ss}), we investigate the different configurations of the three key factors, i.e., representation vectors, removal percentage, and number of eigenvectors.

\noindent\textbf{RQ3: How accurate are the widely-used proxy metrics in reflecting the actual defensive performance?}

\noindent\textbf{Motivation.}
Evaluating the actual defense effectiveness of the SS is expensive and time-consuming because the performance of each defense configuration needs to be verified through a fine-tuning process on a subset of the training data after removing the predicted poisoning data. Existing works~\cite{ramakrishnan2022backdoors} measure Recall, i.e., the percentage of poisons eliminated as a heuristic for the performance of the Spectral Signature in selecting the optimal defense configuration. The reason behind this is that the recall score is cheap to calculate, using only one fine-tuning process. However, it remains unclear whether Recall serves as a reliable proxy for the actual defense effectiveness of SS. The research question aims to investigate the degree of correlation between various proxy metrics, especially Recall
% \hung{@bowen, I want to add "and our proposed metrics", is it okay? or we can introduce later?}
, and the actual defense performance, as quantified by the ASR-D metric, to identify reliable proxies for measuring the performance of SS.

\subsection{Task and Dataset}

% \bowen{please write down our current experimental configuration here}

\textbf{Task.} Following Yang et al.~\cite{ yang2024stealthy}, we use code summarization as the downstream task for code models under our evaluation. Code summarization is a task in which programmers synthesize snippets of code (often code functions, methods, etc.) and summarize the purpose of those snippets in human-readable text. This is a very time-consuming task when programmers need to provide a summarization that is not only accurate to the snippet's function but also follows the summarization instructions so that other programmers can read it more easily. In recent years, this task has been automated by language models \cite{feng2020codebert} \cite{wang2021codet5} to be more accurate and time-efficient for developers. This also raises the vulnerability of language models to backdoor attacks \cite{yang2024stealthy}. 

\textbf{Dataset.}\hung{Trimmed here}
\fix{We use the Python split of CodeSearchNet\cite{husain2019codesearchnet}, a widely used benchmark for code summarization. Following \cite{ramakrishnan2022backdoors,yang2024stealthy}, we use a static attack target: all poisoned examples produce the summarization ``This function is load from the disk safely.'', with their designed trigger.}

% We use the popular public dataset CodeSearchNet; the dataset, while originally used for the code search task, has been used as a benchmark for the code summarization task on many models~\cite{wang2021codet5,feng2020codebert}. The dataset comprises multiple programming languages such as Java, Go, Python, etc. In this task, we used Python-specific code to evaluate the Spectral Signature against attacks. 
% % \textbf{Poison configuration.} 
% Moreover, every poisoning strategy will be executed under a static target. A static target is where the predictions for all attack messages are the same. In this specific scenario, following previous literature \cite{yang2024stealthy}, every attack message will have the same summarization text as \textit{``This function is load from the disk safely.''}
% % \textbf{Spectral Signature baselines.} 
% Besides, we experience different Spectral Signature configurations, current literature ~\cite{ramakrishnan2022backdoors} sets the number of eigenvectors to 10, and the removal ratio to \(1.5\) times the poisoning rate. However, we argue that the poisoning dataset should be agnostic regarding the poisoning rate, so it is unrealistic to indicate the proportion of removal based on the poisoning rate. Therefore, we follow the removal percentage in this paper to be \(10\%\) and \(15\%\).

\subsection{Models}\label{sec:models}

% Motivated by the success of Transformer-based pre-trained models in natural language processing, such as BERT~\cite{devlin2019bert} and RoBERTa~\cite{liu2019roberta}, recent years have witnessed the widespread development of pre-trained models for source code. These models have demonstrated state-of-the-art performance across a wide range of software engineering tasks, such as code generation~\cite{wang2021codet5, le2022coderl}, code comprehension~\cite{feng2020codebert, chapagain2025automated}, and program analysis~\cite{le2022autopruner, ye2022neural}. Given their popularity and potential, this study focuses on two well-known pre-trained code models: CodeBERT~\cite{feng2020codebert} and CodeT5~\cite{wang2021codet5}.

Our main experiment (Section~\ref{sec:results}) primarily focuses on the two widely-used models in code backdoor attack literature~\cite{yang2024stealthy,sun2023backdooring,li2023multi,wan2022you,li2022poison}, CodeBERT~\cite{feng2020codebert} and CodeT5~\cite{wang2021codet5}.
CodeBERT is a bimodal pre-trained model for programming languages (PL) and natural language (NL).  The CodeBERT architecture is Transformer-based and is exactly like RoBERTa-base with an encoder-only architecture. CodeBERT has 12 transformer blocks with 12 attention heads. CodeBERT has been used to complete many tasks related to code, such as code search (NL to PL), code summarization (PL to NL), code completion, or code translation.
CodeT5 is one of the earliest models that uses an encoder-decoder architecture for coding tasks. The architecture of both the encoder and decoder of CodeT5 contains 12 transformer blocks. CodeT5, in addition to the downstream tasks from CodeBERT, has been trained for code generation, clone detection, code refinement, and defect detection.
We further expand our experiment by considering the more recent model \textbf{Qwen2.5-Coder} in the Discussion section (Section ~\ref{subsec:qwen-model-exp}).

% \hung{\textbf{Qwen2.5-Coder}The architecture of Qwen2.5-Coder is  a decoder-only architecture, derived directly from Qwen2.5, focus on coding task. Qwen2.5-Coder came with variety of model sizes, from 0.5B to 32B parameters, was a State-of-the-art Code LLM model at the time it released, match the coding ability with GPT-4o.}

\subsection{Evaluation Metrics}~\label{sec:eval_metrics}

\subsubsection{Evaluation on the detected poisoning instances} To measure the effectiveness of the defense method in each configuration based on the  number of poison code examples, we will use the following evaluation metrics: 

% \bowen{for each of the metrics described below, please clarify at the end of its description, what is the range of its values? [0,1]? is it the higher the better, or the lower value the better?}

\begin{itemize}[wide=0pt]
    
    \item \textbf{Recall}: measures the percentage of poisoned samples correctly predicted by the defense method. It has been widely adopted in prior studies as a proxy for evaluating the effectiveness of backdoor defense techniques~\cite{wan2022you,sun2023backdooring,sun2022coprotector,schuster2021you}. Range from 0 (0\%) to 1 (100\%), a higher recall value leads to a higher number of poison detected
      \begin{equation}
    Recall = \frac{\text{\# Correctly Predicted Poisoned Samples}}{\text{\# Actual Poisoned Samples}}
  \end{equation}
\item \textbf{False Positive Rate (FPR)}: measures the proportion of clean examples that has been detected as poison over the total number of predicted poison examples \cite{wan2022you, sun2023backdooring,ramakrishnan2022backdoors,yang2024stealthy}. The metric value ranged from 0 (0\%) to 1 (100\%), A high FPR indicates the defense method has made more mistake when detects more clean samples as poison ones.
% \bowen{if I am not wrong, FPR is widely used. Pls cite more papers.}

\begin{equation}
    FPR=\frac{\text{\# Clean Samples Predicted Incorrectly}}{\# \text{Samples That Predicted Poison}}
\end{equation}
% \fix{C1.7}{\item \textbf{False Positive Rate (FPR)}: measures the proportion of clean examples that has been detected as poison over total number of predicted poison examples \cite{wan2022you, sun2023backdooring,ramakrishnan2022backdoors,yang2024stealthy}. The metric value ranged from 0 (0\%) to 1 (100\%), high FPR indicates the defense method has made more mistake when detects more clean samples as poison ones.
% % \bowen{if I am not wrong, FPR is widely used. Pls cite more papers.}

% \begin{equation}
%     FPR=\frac{\text{\# Clean Samples Predicted Incorrectly}}{\# \text{Samples That Predicted Poison}}
% \end{equation}
% }

% \item \textbf{Precision}: measure the percentage of correct poison detected over total number of poison predictions. 

% \fix{C1.7}{Sun et al.\cite{sun2023backdooring} proposed FPR (FULLNAME HERE), measured by calculate proportion of clean examples that has been detected as poison over total number of predicted poison examples. However, for standardlise the metrics, we measure precision, which measures the percentage of correct poison detected over total number of poison predictions. Precision and FPR are similar since the sum of them is 1.}
% \bowen{1. we need to clarify what does FPR mean first, 2. do you mean the defintion of FRP is same as precision? Pls clarify this.}
% \bowen{as discussed, pls use FPR instead of precision.}

% \begin{equation}
%     Precision=\frac{\text{\# Correctly Predicted Poison Samples}}{\# \text{Samples that predicted poison}}
% \end{equation}

\item \textbf{Negative Predictive Value (NPV):} We proposed the Negative Predictive Value (NPV) measures correctly predicted clean samples over the total samples predicted clean. NPV ranges from 0 (0\%) to 1 (100\%), a higher NPV value in backdoor defense scenario means a higher portion of remaining dataset is clean.
\begin{equation}
    NPV= \frac{\text{\# Correctly predicted Clean Samples}}{\text{\# Predicted Clean Samples}}
\end{equation}

  \end{itemize}

\subsubsection{Evaluation on attack performance}\label{sec:eval-metrics-attack} To measure the effectiveness of the attack method or defense method on inference, we use following metrics:
  
  \begin{itemize}[wide=0pt]

  \item \textbf{Attack Success Rate ASR ~\cite{ramakrishnan2022backdoors}} measure the effectiveness of the attack message directly. ASR is computed from the output of test dataset, where we compute total number of poison data that predict poisoned target correctly over total number of poison data from test dataset. ASR ranged from 0(\%) to 100(\%), higher ASR mean a more dangerous attack method when a sample is poisoned.
  \begin{equation}
    ASR = \frac{\text{\# Poison Samples predicted correctly}}{\text{\# Poison Samples}}
  \end{equation}
  
  % \item \textbf{False Positive Rate~\cite{yang2024stealthy,wan2022you,sun2022coprotector,schuster2021you,li2024poison}}: This is the percentage of normal (unpoisoned) samples that were incorrectly identified as poisoned by the defense method.
  % \begin{equation}
  %   FPR = \frac{\text{\# Actual Clean Samples mispredicted by defense method}}{\text{\# Clean Samples}}
  % \end{equation}
  \item \textbf{Attack Success Rate Under Defense~\cite{yang2024stealthy}}, denoted by ASR-D. Note that, a certain value of \textit{Recall} does not always guarantee to inactivate the triggers. Therefore, we will use ASR-D introduced in~\cite{yang2024stealthy}.
\fix{Similar to ASR, ASR-D calculated ASR from remain dataset. The  range of ASR-D is from 0(\%) to 100(\%), a higher ASR-D means a more dangerous attack method when a sample is poisoned even after defense.} \hung{reduced here}
  % ASR-D is used to measure the attack performance when the defense is used to detect poisoned examples. To protect the model from backdoor attacks, we apply defense methods to both the training and test data. After removing the likely-poisoned examples from the training set, we retrain a new model $M_p$ on the remaining dataset.
  % On the test dataset, we only feed the examples that are not labeled as likely-poisoned examples to the model. 
  % We define ASR-D as follows.
  
    % ASR-D =  \frac{\sum_{x_i \in \mathcal{X}} M_p(x_i) = \tau \land \neg \mathcal{S}(x_i) }{\sum_{x_i \in \mathcal{X}}  x_i~\text{contains triggers}}
  % \begin{equation}

  %   $$ASR-D = \frac{\text{\# }}{\text {\# }}$$
    
  % \end{equation}
  \begin{equation}
    \textit{ASR-D} = \frac{\text{\# Remain Poison Samples predicted correctly}}{\text{\# Remain Poison Samples}}
  \end{equation}
    % If $\mathcal{S}(x_i)$ is true, it means that the example $x_i$ is detected as poisoned example. $\sum_{x_i \in \mathcal{X}} M_p(x_i) = \tau \land \neg \mathcal{S}(x_i)$ means the number of all the poisoned examples that are not detected by defense methods and produce success attacks. 
\end{itemize}

\subsubsection{Evaluation on the model performance for downstream tasks}
To assess the effectiveness of the evaluated models on the downstream task of code summarization, we employ the BLEU score, a widely recognized and frequently used evaluation metric, which is described in detail below.

\begin{itemize}[wide=0pt]
    \item \textbf{BLEU\cite{papineni2002bleu}}: 
    % This metric is a common metric measure the similarity of 2 corpus, typically used to evaluate quality of machine-generated translations by comparing them to human translations. BLEU score is measured by means of n-gram precision \(p_n\) with \(n\) span from \(1\) to \(N\), followed by penalty score from reference length \(r\) and prediction length \(c\). In code summarization, BLEU measures the model performance of the language model. The range of BLEU is from 0 to 100, where a higher BLEU shows a more accurate summarization in our study.
    \fix{This is a common metric to measure the similarity of 2 corpus, typically used to evaluate the quality of machine-generated translations compared to human benchmark. The range of BLEU is from 0 to 100, where a higher BLEU  is better.}\hung{trimed here}
      \begin{equation}
        BLEU = min(exp(1-\frac{r}{c}),1)*exp(\Sigma_{n=1}^{N}{\frac{1}{N}log(p_n)})  
  \end{equation}
\end{itemize}

\subsubsection{Evaluation on correlation between 2 variables} To demonstrate the relationship between a heuristic and actual results, we utilize two widely used rank-order correlation coefficients: Spearman's and Kendall $ \tau$'s,  \fix{non-parametric rank correlation coefficients ranging from -1 to 1, where values near ±1 indicate strong monotonic correlation.}\hung{trimmed here}
    % \textbf{Spearman} is define as Pearson correlation coefficient\cite{bravais1844analyse} between the rank of variables.

    \begin{equation}
        Spearman=\rho [R[X],R[Y]]=\frac{cov[R[X],R[Y]]}{\sigma_{R[X]}\sigma_{R[Y]}}
    \end{equation}
    
    % \textbf{Kendall $\tau$}\cite{kendall1938new} measure association between variables using order between pairs of values. The statistic use concordant and discordant of 2 pairs of value, where the sort of order between 2 variables agree.
    \begin{equation}
        Kendall-\tau=\frac{(\text{\#concordant pairs})-(\text{\#discordant pairs})}{\text{\# total pairs}}
    \end{equation}

\section{Results}\label{sec:results}

% \bowen{@hung, add the missing refs}

\subsection*{\textbf{RQ1: Is the usage of Spectral Signature in the existing literature optimal for code backdoor detection?}}

% \hung{Each paragraph describe one thing clearly}

\begin{table}[h]
\centering
\caption{Comparison of ASR-D and Task Performance Differences (Original and Cleaned after Poisoned) for Best vs. Used Setting Across Different Attack Methods and Poison Rates}
\label{tab:current_vs_optimal}
\resizebox{\columnwidth}{!}{
\begin{tabular}{@{}llccc@{\hspace{1em}}ccc@{}}
\toprule
\textbf{Attack} & \textbf{Rate} 
& \multicolumn{3}{c@{\hspace{1em}}}{\textbf{Best Setting}} 
& \multicolumn{3}{c}{\textbf{Used Setting}} \\
\cmidrule(lr){3-5} \cmidrule(lr){6-8}
& & \textbf{ASR-D} & \multicolumn{2}{c}{$\Delta$ BLEU} & \textbf{ASR-D} & \multicolumn{2}{c}{$\Delta$ BLEU} \\
\cmidrule(lr){4-5} \cmidrule(lr){7-8}
& & & \textbf{Original} & \textbf{Cleaned} & & \textbf{Original} & \textbf{Cleaned} \\
\midrule

\multirow{3}{*}{\textbf{Fixed}} 
& 1\% & 0.00\% & -0.21 & -0.21 & \textbf{0.00\%} & -0.21 & -0.21 \\
& 5\% & \textbf{28.07\%} & -0.15 & 0.22 & 98.25\% & -0.18 & 0.17 \\
& 10\% & \textbf{42.11\%} & -0.98 & -0.36 & 100.00\% & 0.09 & 2.40 \\

\midrule

\multirow{3}{*}{\textbf{Grammatical}} 
& 1\% & 0.00\% & 0.57 & 0.57 & \textbf{0.00\%} & 0.57 & 0.57 \\
& 5\% & \textbf{4.29\%} & -1.03 & -0.57 & 100.00\% & -0.03 & 0.46 \\
& 10\% & \textbf{99.02\%} & -0.44 & 0.21 & 100.00\% & 0.12 & 2.39 \\

\midrule

\multirow{3}{*}{\textbf{Adaptive}} 
& 1\% & \textbf{0.00\%} & -0.16 & -0.16 & 0.00\% & 0.73 & 0.73 \\
& 5\% & 0.00\% & 0.04 & 0.04 & \textbf{0.00\%} & 0.04 & 0.04 \\
& 10\% & \textbf{13.68\%} & 0.00 & 0.60 & 38.95\% & 0.13 & 0.74 \\

\bottomrule
\end{tabular}
}
\\\vspace{1mm}
*: \textbf{Bolded number} represent lower ASR-D between Used and Best Setting
% \vspace{-2mm}
\end{table}

% \begin{figure}[]
%   \centering
%   \includegraphics[width=0.8\linewidth]{imgs/compare_line_encoder_decoder.png}
%   \caption{Accuracy score comparison between decoder and encoder}
%   \label{graph:acc_decoder_encoder}
% \end{figure}

\subsubsection*{\textbf{Experimental setting}}

\textbf{Attack configurations.} 
For each configuration of SS within the configuration space (described in Section~\ref{sec:config-space}), we evaluate its performance over three popular types of attacks (\texttt{Fixed}, \texttt{Grammatical}, and \texttt{Adaptive}) with three commonly used poisoning rates (1\%, 5\%, 10\%).
As a result, we evaluated 252 (28$\times$3$\times$3) experimental settings.
This enables a comprehensive analysis of SS behavior, thereby improving the generalizability of our findings.
% This enables a robust analysis of the impact of each configuration on SS performance.
Following previous work \cite{ramakrishnan2022backdoors,yang2024stealthy}, we use ASR-D (described in Section~\ref{sec:eval-metrics-attack}) as the main evaluation metric to measure the effectiveness of an SS configuration.
% Each configuration produces a new filtered dataset that will be re-trained to obtain new test results, and then we calculate ASR-D based on those new test results. However, due to limitations in computing resources, the list of the number of eigenvectors (k) has been compacted to (1,2,3,10,15,20,50). This selection is based on: (1) the numbers included provide the best recall score in one or a few configurations, and (2) these numbers cover a small (1, 2, 3), medium (10, 15, 20) and large (50) number of k.
\textbf{SS configuration used in existing works.}
% \bowen{RQ1 is about \textbf{any configuration in our defined configuration space} vs. \textbf{configuration used in the existing works}, so we also need to describe what are the SS configuration used in existing works.}
Following the initial adaptation of Ramakrishnan et al., subsequent studies~\cite{yang2024stealthy, sun2023backdooring} commonly employ a configuration with $k=10$ and a removal rate of 1.5 times the poisoning rate. We refer to this configuration as ``used configuration''.

\subsubsection*{\textbf{Results}}
% Based on the aforementioned experimental configurations, we conduct an exhaustive evaluation of the Spectral Signature (SS) under various configurations to identify optimal defense parameters.
% \bowen{why we present two tables based on CodeBERT and CodeT5 separately here? it does not matter in RQ1 as long as any configuration is better than the used one.}

% \bowen{pls describe in the following flow:
% \begin{itemize}
%     \item Paragraph 1: Table~\ref{} presents the comparison between our identified best configuration and the used configuration. We found that, among 9 attack configurations, we found that in XX of them, the used configurations are not optimal. On averaged, the best configurations we identified outperform the used configurations by YY\%.
%     % \item Paragraph 2: Moreover, we found that, among all the ?? attack configurations, we found that the majority of the best configuration are based on K=??, model=??, removal ratio=??.
%     \item Paragraph 2: Besides, we also found that, with the identified best configuration, SS's performance can be significantly better than the used configuration for the more advanced attack, i.e., adaptive trigger-based. (provide concrete numbers to support this statement)
% \end{itemize}
% }

% \bowen{pls don't use textit{} and texttt format. instead, for trigger type, pls use the capitalized case, i.e., Fixed, Grammatical, Adaptive.}\hung{fixed}

Table~\ref{tab:current_vs_optimal} presents a comparison between our identified best configuration and the used configuration. The table presents three evaluation metrics on each attack configuration: ASR-D, BLEU of unpoisoned dataset, and \textit{BLEU} of the poisoned dataset after being cleaned by SS.
We found that, among the nine attack configurations, in six of them, the used configurations are not optimal. On average, the best configurations we identified outperform the used configuration, reducing the ASR-D of attackers by 41.67\%. 

Besides, we also found that, with the identified best configuration, SS's performance can be  better than the configuration used under higher poison rate attacks (5\% and 10\%). On \texttt{Fixed} trigger at 5\% and 10\%, the optimal configuration reduces 70.18\% and 57.89\% ASR-D compared to the used configuration, respectively. In the case of a 5\% \texttt{Grammatical} trigger, it reduced 95.71\% ASR-D compared to the used configuration. For a 10\% poison rate of \texttt{Adaptive} trigger, it has a  25.27\% ASR-D score lower than the used configuration.

\begin{summarybox}[title={Answer summary to RQ1}]
The usage of Spectral Signature in the existing literature is not optimal, especially with more recent attack methods.
The average improvement of the optimal configuration from the previously recommended configuration is around a 40\% reduction on average.
% Using CodeT5 for Spectral Signature is likely a better defense against Adaptive trigger, while using CodeBERT shows good defense against Fix and Grammatical triggers. 
\end{summarybox}

\subsection*{\textbf{RQ2: How does each factor impact SS's performance?}}\label{subsec:RQ2_result}

% \begin{figure*}[]
%   \centering
%   \includegraphics[width=\linewidth]{imgs/RQ2_CodeBERT.png}
%   \caption{The change of ASR-D based on different values of \textit{k} and removal percentage with CodeBERT representation}
%   % \bowen{ASR in subfigs should be ASR-D?}
%   % \hung{fixed}
%   \vspace{-5mm}
%   \label{graph2.1:asr_CodeBERT_3d}
% \end{figure*}

% \begin{figure*}[]
%   \centering
%   \includegraphics[width=\linewidth]{imgs/RQ2_codet5.png}
%   \caption{The change of ASR-D based on different values of \textit{k} and removal percentage with CodeT5 representation}
%   % \bowen{ASR in subfigs should be ASR-D?}
%   \label{graph2.1:asr_CodeT5_3d}
% \end{figure*}

\subsubsection*{\textbf{Experimental settings}}

To answer RQ2, we collect and analyze the results of SS with all configurations from the experiments in RQ1. Particularly, we measure the impact of the number of eigenvectors ($k$), removal percentage, and code representation models on SS's performance against different attack scenarios. The analysis focuses on how these parameters influence the effectiveness of SS, as quantified by the metrics, ASR-D \fix{and BLEU score}. Table~\ref{RQ2:k_value}, ~\ref{RQ2:models}, and~\ref{RQ2:removal rate} provide 
\fix{the effects of different Spectral Signature parameters, including $k$ values, representation model, and removal percentage,}
% \hung{the effect of different aspects of Spectral Signature, including k values, model, and removal percentage,} 
across the full spectrum of configurations and attack scenarios, respectively.
\fix{For $k$ values, for simplicity and convenience in subsequent analysis, we separate them into two groups: $k=[1,2,3]$ and $k=[10,15,20,50]$, referred to as low $k$ and high $k$, respectively. This division is based on the finding that high $k$ ($k=10$) outperforms low $k$ ($k=1$) in Code models ~\cite{ramakrishnan2022backdoors}, along with our intention to explore values beyond $k=10$.}
% \bowen{I don't understand the following sentence, what does 95\% confidence interval here mean? Usually we say 95\% confidence interval only when we report statistic significance, i.e., p-value}\hung{My intention is that in some aspects, for example k>10 vs k<10 in 1\%, the k>10 for asr-d is significantly higher than k<10.}
\fix{For each group, we report the mean value and 95\% confidence interval of the BLEU score and ASR-D, separated by the $\pm$ sign. The lower ASR-D score in each column is highlighted.}
% \hung{the effects of different Spectral Signature parameters, including $k$ values, representation model, and removal percentage,}
% \hung{For each group, we report mean value and 95\% confidence interval of BLEU score and ASR-D, separated by ± sign. A lower  ASR-D score of each column will be highlighted.}
% with the result of RQ1, we collect the performance of all the configurations of SS within the configuration space.
% Then, we visualize and analyze the results on the impact factor basis.
% For clarity in the subsequent analysis, we define \textit{recall@k} and \textit{ASR-D@k} as the recall and ASR-D, respectively, evaluated at a defense configuration using the top-$k$ eigenvectors.

% showed the ASR-D results of all defense configurations against each attack method and poisoning rate. Prior to analysis, we denote \textit{recall@k} or \textit{ASR-D@k} as values of recall, ASR-D at \textit{k} number of eigenvectors in defense configurations. Note that an optimal defense or defense configuration wants to achieve \textit{ASR} score as low as possible. All of the graphs in Figure.\ref{graph2.1:asr_CodeBERT_3d} and \ref{graph2.1:asr_CodeT5_3d}, each representing a distinct attack configuration, represent the influence of removal rates and number of eigenvectors on ASR-D. 

\subsubsection*{\textbf{Result}}

\begin{table}[H]
  \centering
  
  \caption{Effect of different k value groups \\ on different poison rates}
  % \bowen{here in the caption, we say different k values, but in the table, we only present k $<$ 10 and k $\ge$10, why?}
\begin{tabular}{ccccccc}
\toprule
\textbf{\begin{tabular}[c]{@{}c@{}}Poison\\ percen-\\tage\end{tabular}} & \multicolumn{2}{c}{\textbf{1\%}}                                                                                         & \multicolumn{2}{c}{\textbf{5\%}}                                                                                          & \multicolumn{2}{c}{\textbf{10\%}}                                                                                        \\ \midrule
\textbf{\begin{tabular}[c]{@{}c@{}}number\\of\\eigen-\\vector\end{tabular}}                                                       & \textbf{BLEU}                                         & \textbf{ASR-D}                                                  & \textbf{BLEU}                                         & \textbf{ASR-D}                                                   & \textbf{BLEU}                                         & \textbf{ASR-D}                                                  \\ \midrule
\textbf{k $<$ 10}                                                        & \begin{tabular}[c]{@{}c@{}}18.74\\ ±0.24\end{tabular} & \begin{tabular}[c]{@{}c@{}}97.80\%\\ ±3.24\end{tabular}          & \begin{tabular}[c]{@{}c@{}}18.08\\ ±0.41\end{tabular} & \textbf{\begin{tabular}[c]{@{}c@{}}75.86\%\\ ±11.70\end{tabular}} & \begin{tabular}[c]{@{}c@{}}18.09\\ ±0.13\end{tabular} & \textbf{\begin{tabular}[c]{@{}c@{}}90.14\%\\ ±7.86\end{tabular}} \\ \midrule
\textbf{k $\ge$10}                                                       & \begin{tabular}[c]{@{}c@{}}17.84\\ ±0.14\end{tabular} & \textbf{\begin{tabular}[c]{@{}c@{}}10.80\%\\ ±6.78\end{tabular}} & \begin{tabular}[c]{@{}c@{}}17.95\\ ±0.12\end{tabular} & \begin{tabular}[c]{@{}c@{}}82.23\%\\ ±10.38\end{tabular}          & \begin{tabular}[c]{@{}c@{}}19.00\\ ±0.62\end{tabular} & \begin{tabular}[c]{@{}c@{}}91.65\%\\ ±6.35\end{tabular}\\ \bottomrule
\end{tabular}
\label{RQ2:k_value}
% \vspace{-3mm}
\end{table}

\begin{table}[H]
  \centering
  
  \caption{Effect of different representation models \\ on different attack methods} 
  % \bowen{this might be confusing as we mention qwen models in section~\ref{sec:models}  }
  
\label{RQ2:models}

\begin{tabular}{ccccccc}
\toprule
\textbf{Attack}                                              & \multicolumn{2}{c}{\textbf{Adaptive}}                                                                                     & \multicolumn{2}{c}{\textbf{Fixed}}                                                                                         & \multicolumn{2}{c}{\textbf{Grammatical}}                                                                                   \\ \midrule
\textbf{Model}                                               & \textbf{BLEU}                                         & \textbf{ASR-D}                                                    & \textbf{BLEU}                                           & \textbf{ASR-D}                                                   & \textbf{BLEU}                                          & \textbf{ASR-D}                                                    \\ \midrule
\textbf{\begin{tabular}[c]{@{}c@{}}Code\\ BERT\end{tabular}} & \begin{tabular}[c]{@{}c@{}}17.78\\ ±0.39\end{tabular} & \begin{tabular}[c]{@{}c@{}}81.80\%\\ ±20.14\end{tabular}          & \begin{tabular}[c]{@{}c@{}}17.82\\ ±0.15\end{tabular}   & \textbf{\begin{tabular}[c]{@{}c@{}}95.61\%\\ ±8.60\end{tabular}} & \begin{tabular}[c]{@{}c@{}}17.66\\ ±0.19\end{tabular}  & \textbf{\begin{tabular}[c]{@{}c@{}}86.25\%\\ ±21.58\end{tabular}} \\ \midrule
\textbf{\begin{tabular}[c]{@{}c@{}}Code\\ T5\end{tabular}}                                              & \begin{tabular}[c]{@{}c@{}}18.76\\ ±0.39\end{tabular} & \textbf{\begin{tabular}[c]{@{}c@{}}71.11\%\\ ±28.56\end{tabular}} & \begin{tabular}[c]{@{}c@{}}18.83\\ ±0.42\end{tabular} & \begin{tabular}[c]{@{}c@{}}100.00\%\\ ±0.00\end{tabular}           & \begin{tabular}[c]{@{}c@{}}18.83\\ ±0.38\end{tabular} & \begin{tabular}[c]{@{}c@{}}99.64\%\\ ±0.70\end{tabular}\\ \bottomrule           
\end{tabular}
\vspace{-3mm}
\end{table}
% \subsubsection{Experiemental configuration} 

\begin{table}[H]
  \centering
  \caption{Effect of different removal percentage \\ on different attack methods}
\begin{tabular}{ccc}
\toprule
\textbf{\begin{tabular}[c]{@{}c@{}}Removal\\  percentage\end{tabular}} & \textbf{BLEU} & \textbf{ASR-D} \\ \midrule
\textbf{10\%} & \begin{tabular}[c]{@{}c@{}}18.40\\ ±0.23\end{tabular} & \begin{tabular}[c]{@{}c@{}}78.44\%
\\ ±6.73\end{tabular} \\ \midrule
\textbf{15\%} & \begin{tabular}[c]{@{}c@{}}18.16\\ ±0.20\end{tabular} & \textbf{\begin{tabular}[c]{@{}c@{}}67.29\%\\ ±7.47\end{tabular}}\\
\bottomrule
\end{tabular}
\label{RQ2:removal rate}
\end{table}

\noindent\textbf{Impact of Number of Eigenvectors (\textit{K})}: 
%\bowen{@hung, pls discuss here on the impact of different K here}
% \bowen{at the beginning here, can you specify which table?}
\fix{Table \ref{RQ2:k_value}} showed that the impact of \textit{k} is highly influenced by the poisoning rate. At a 1\% poisoning rate, all attack methods are successfully mitigated by defense configurations with a higher number of eigenvectors \textit{k}. Specifically, for configurations where \(k>=10\), where average \textit{ASR-D@k} is 10.8\%, indicating a high rate of defense success.
% \bowen{I am lost in the previous sentence, in Table V, I do not see 0\%.}\hung{changed to ignore specific case.}
In contrast, at the same poison rate, with configurations of $k<10$, SS  does not defend well against any attack methods, with \textit{ASR-D@k} mostly at 100\%. A higher poisoning rate, on the other hand, shows the opposite trend. At 5\% and 10\%, it requires a small \textit{k} to defend effectively, \textit{ASR-D@k} with $k<10$ is, \fix{on average, 6.36\% and 1.51\% lower than that of $k\geq10$, respectively.}
% \hung{on average, has 6.36\% and 1.51\% drop compare to $k\geq10$, respectively.  }

%on average, only 38.62\% for CodeT5 and 57.64\% for CodeBERT. At a 10\% poisoning rate, although the spectral signature at all configurations is not as good in defense, a lower number of eigenvectors also indicates a good impact against the attack. The average \textit{ASR-D@2} score is 73.6\% for CodeBERT and 71.2\% for CodeT5, which are the lowest among other \textit{k}.

\noindent\textbf{Impact of \textit{representation model}}: 
%\bowen{@hung, pls discuss here on the impact of different models here}
Table \ref{RQ2:models} shows that the representation vector of CodeBERT and CodeT5 for SS is effective against different types of attacks. We found that using CodeBERT for Spectral Signature gives better performance in \texttt{Fixed} and \texttt{Grammatical}, while CodeT5 has been shown to be better at preventing the \texttt{Adaptive} trigger. 
\fix{Against \texttt{Fixed} and \texttt{Grammatical} triggers, SS using CodeBERT achieves, on average, 5\% and 14\% lower ASR-D than that of CodeT5, respectively. For instance, at a 10\% poison rate with the \texttt{Fixed} trigger, CodeBERT's representation achieves 42.11\% ASR-D while CodeT5's yields 100\% ASR-D. On the other hand, CodeT5, on average, achieves 10.69\% lower ASR-D compared to CodeBERT when defending against the \texttt{Adaptive} trigger.}
% \hung{Against Fixed and Grammatical triggers,  SS using CodeBERT achieves, on average, 5\% and 14\% lower ASR-D than that of CodeT5, respectivly. For instance, at a 10\% poison rate of Fixed trigger, CodeBERT's representation achieves 42.11\% ASR-D while CodeT5's has 100\% ASR-D. On the other hand, CodeT5, on average, achieves 10.69\% lower ASR-D compared to CodeBERT when defending against Adaptive trigger.}

\noindent\textbf{Impact of \textit{removal percentage}}: 
%\bowen{@hung, pls discuss here on the impact of different models here}
% \bowen{at the beginning here, can you specify which table?}
% \bowen{what figure ur talking about the in next sentence?}\fix{\textbf{(adjusted)}} 
\fix{Table \ref{RQ2:removal rate}} indicates that a higher removal percentage often leads to a more successful defense. Higher removal percentage is directly increase the number of poison data detected. On average, ASR-D of 15\% removal percentage achieves 24.45\% lower than that of 10\%. 
\fix{However, in some scenarios, the 10\% removal percentage also outperforms the 15\% removal percentage. This can be explained by RQ3, where we discuss the existence of NPV and its influence on Spectral Signature performance.}

\fix{\noindent \textbf{Impact on \textit{BLEU}:} Overall, the individual factors of Spectral Signature do not have a major effect on BLEU performance. Across all aspects, the performance of cleaned models does not exceed a 1-point difference compared to the original model performance.}

\begin{summarybox}[title={Answer summary to RQ2}]
In general, there is no golden configuration of Spectral Signature that can be universally effective across all attack scenarios. However, our empirical observations reveal a consistent pattern: a higher number of eigenvectors (\textit{k}) is best for defending against lower poisoning rates, whereas smaller \textit{k} values are more effective at mitigating higher poisoning rates.

% \fix{C1.8 new}{We advocate future research on estimating the poisoning rate to reach generalization of spectral signature defense strategy}

% \fix{C1.8}{In general, there is no golden configuration of Spectral Signature that can be universally effective across all attack scenarios. However, previous attack method only apply low poison rate percentage  (Wan et al.\cite{wan2022you} use 5\% and 1\% for their fix/Grammatical triggers, and Zhang et al.\cite{zhang2021advdoor} use 0.5\%, 1\% and 5\% in AFARAIDOOR) for the stealthiness purpose. Therefore, it is recommended to use a high number of eigenvectors to apply Spectral Signature for defense purpose.}

% \bowen{I think we can keep the old version of the answer summary but add one more sentence at the end like "We advocate future research on estimating the poisoning rate to improve the effectiveness of spectral signature."}

\end{summarybox}

\subsection*{\textbf{RQ3: How accurate are the widely-used proxy metrics in reflecting the actual defensive performance?}}\label{sec:RQ3_result}

% \bowen{configuration: adv ----> calculate the recall and ASR-D}

\subsubsection{Experimental settings}
To answer RQ3, we collect the values of ASR-D alongside a set of commonly used proxy metrics that serve as estimators of the actual performance of a given configuration, as presented in Section~\ref{sec:eval_metrics}. 
% \bowen{cite the original paper of each correlation metric in the following sentence}\hung{fixed}
We choose to use two widely used correlations, i.e., Spearman\cite{bravais1844analyse} and Kendall $\tau$~\cite{kendall1945treatment}, because both of them are non-parametric (which means they don't assume a specific data distribution) and can be used to determine the correlation between two ranks (in our case, two ranks based on two different metrics).
Both correlation coefficients are numbers ranging from -1 to 1. Values close to -1 or 1 indicate strong correlation, while values close to 0 indicate weak correlation.

Note that some modifications in these experiments are made for a controlled experiment. First, by definition, a lower value of ASR-D indicates better performance of the defense tool. On the other hand, the greater value of the evaluation metrics like Recall and NPV indicates the better performance of the defensive method. Hence, we will calculate the correlation of these metrics to 1$-$ASR-D.
% \bowen{I don't think we need to say "1-" here as typically, we say positively or negatively correlated. or poistive correlation or negative correlation.} \hung{My result will be reversed at every cells, I can do that but it is not very pleasing to the eyes to see a lot of negative sign so I don't do that, or I could change to -ASR-D} 
% \bowen{sure, i change back to 1-asr-d as it looks less weird than -asrd}
Second, to comprehensively evaluate the effectiveness of each metric, we calculate correlation in every considered configuration and aggregate the results in four core dimensions,
% . We define \textbf{aspects} as factors that contribute to the attack or defense of backdoor attack configurations, in this case, which consist of 
models, poison rate, attack methods, and removal percentage. 
For each dimension, we group the configurations accordingly. For example, the "Removal Percentage" aspect divides the configuration space into 2 groups, 10\% and 15\%, each of which contains a subset of configurations that has the removal percentage 10\% or 15\%. For each group, we calculate the correlation on the whole population of the group.

% \bowen{in the table, we say 10\% and 15\% are poisoning rates, but in previous sentences, we call them removal ratios. pls CAREFULLY proofread to reduce these minor mistakes.}\hung{fixed}
% % \fix{Q1.5 C1.17}{

% }

% investigate the relation of evaluation metrics to the finding of the optimal defense configuration when study Spectral Signature to answer if recall is a good metric to produce good ASR-D. First, similar to RQ2, we calculate different evaluation metric scores on ranges of configuration of Spectral Signature against different configurations. Then, using the \textit{ASD-D} results from RQ2, calculate the correlation of each metric to \textit{ASD-D} score to see what if recall is a good estimation to evaluate the effectiveness of defense configuration.

% In this research question, we want to answered if the defense recall is a good approximation of Spectral Signature effectiveness. To extend the evaluation metrics, we also add precision, accuracy and F1-score as heuristic for select effective configuration. Then, we calculate correlation of each metric to ASR-D using rank-based correlation calculation: Kendall $\tau$ and Spearman correlation. 

\subsubsection{Result}

Table~\ref{tab:correlation_asrd_aspects} presents the results of the correlation coefficients between ASR-D and proxy metrics in different groups; the highest correlation score is highlighted in bold. Overall, we find that FPR does not correlate with ASR-D, while NPV shows the highest correlation with ASR-D. In all cases, NPV consistently shows the highest absolute correlation in every group (0.569-0.871 for Spearman, 0.448-0.727 for Kendall $\tau$). In terms of FPR, its correlation scores fluctuated between positive and negative, and the absolute value doesn't exceed 0.2 in most groups at both Spearman and Kendall $\tau$. While having a similar correlation score compared to NPV with a difference no more than 10\% at some groups, such as at 1\% poison rate, \texttt{Fixed} and \texttt{Adaptive} trigger, Recall correlation score in other groups show a big deficit between it and NPV, from 0.254 to 0.689 at Spearman, and from 0.201 to 0.571 at Kendall $\tau$.
% \bowen{fix [XXX] with specific group name}\hung{fixed}
Additionally, in some groups (such as 10\% and 5\% poison rate groups), the correlation of Recall to ASR-D is near 0 in both correlation types.

\begin{summarybox}[title={Answer summary to RQ3}]
Our results show that, although the widely-used proxy metric, e.g., Recall, shows positive correlation coefficients with ASR-D in most groups. In general, NPV carries roughly 2.5 times higher correlation coefficients than Recall at Spearman and Kendall $\tau$. This result advocates future work to use NPV as the proxy metric of ASR-D.
\end{summarybox}
\begin{table}[h]
\centering
\renewcommand{\arraystretch}{1.05}
\caption{Correlation Coefficients Between ASR-D and Proxy Metrics}
\label{tab:correlation_asrd_aspects}
\begin{tabular}{p{1cm}ccccc}
\toprule
\textbf{\begin{tabular}{cc}
       Corre-\\lation \\
      Type
\end{tabular}} & \textbf{\begin{tabular}{cc}
     Aspect\\ Name 
\end{tabular}} & \textbf{Group Name} & \textbf{Recall} & \textbf{FPR} & \textbf{NPV} \\ \midrule
\multirow{11}{*}{Spearman} & \multirow{2}{*}{\begin{tabular}{ll}
     
    Removal \\ Percentage
\end{tabular}} & 10\% & 0.449 & 0.141 & \textbf{0.703} \\ %\cline{3-6} 
 &  & 15\% & 0.122 & 0.085 & \textbf{0.811} \\ \cline{2-6} 
 & \multirow{2}{*}{Model} & CodeBERT & 0.302 & 0.197 & \textbf{0.747} \\ %\cline{3-6} 
 &  & CodeT5 & 0.303 & 0.149 & \textbf{0.763} \\ \cline{2-6} 
 & \multirow{3}{*}{Poison rate} & 0.01 & 0.863 & -0.779 & \textbf{0.871} \\ %\cline{3-6} 
 &  & 0.05 & -0.045 & -0.268 & \textbf{0.569} \\ %\cline{3-6} 
 &  & 0.1 & 0.01 & -0.21 & \textbf{0.568} \\ \cline{2-6} 
 & \multirow{3}{*}{\begin{tabular}{ll}
     
    Attack \\ method
\end{tabular}} & Fixed & 0.689 & 0.38 & \textbf{0.712} \\ %\cline{3-6} 
 &  & Grammatical & 0.108 & 0.436 & \textbf{0.721} \\ %\cline{3-6} 
 &  & Adaptive & 0.828 & -0.136 & \textbf{0.864} \\ \cline{2-6} 
 & \multicolumn{2}{c}{Combined} & 0.311 & 0.169 & \textbf{0.761} \\ \midrule
\multirow{11}{*}{Kendall $\tau$} & \multirow{2}{*}{\begin{tabular}{ll}
     
    Removal \\ Percentage
\end{tabular}} & 10\% & 0.361 & 0.1 & \textbf{0.562} \\ %\cline{3-6} 
 &  & 15\% & 0.103 & 0.054 & \textbf{0.674} \\ \cline{2-6} 
 & \multirow{2}{*}{Model} & CodeBERT & 0.245 & 0.153 & \textbf{0.601} \\ %\cline{3-6} 
 &  & CodeT5 & 0.243 & 0.119 & \textbf{0.609} \\ \cline{2-6} 
 & \multirow{3}{*}{Poison rate} & 0.01 & 0.722 & -0.619 & \textbf{0.727} \\ %\cline{3-6} 
 &  & 0.05 & -0.019 & -0.213 & \textbf{0.459} \\ %\cline{3-6} 
 &  & 0.1 & 0.024 & -0.164 & \textbf{0.448} \\ \cline{2-6} 
 & \multirow{3}{*}{\begin{tabular}{ll}
     
    Attack \\ method
\end{tabular}} & Fixed & 0.592 & 0.332 & \textbf{0.597} \\ %\cline{3-6} 
 &  & Grammatical & 0.085 & 0.352 & \textbf{0.598} \\% \cline{3-6} 
 &  & Adaptive & 0.648 & -0.073 & \textbf{0.687} \\ \cline{2-6} 
 & \multicolumn{2}{c}{Combined} & 0.255 & 0.132 & \textbf{0.614} \\ \bottomrule

\end{tabular}

% \vspace{-2mm}
% \vspace{-0m}
\end{table}
% \begin{figure*}[]
%   \centering
%   \includegraphics[width=0.8\linewidth]{imgs/recall/RQ3.png}
%   \caption{recall graph in 1\% poisoning rate on every number of eigenvectors}
%   \label{graph5.5:Recall_all_attack}
% \end{figure*}

\section{Discussion}

\subsection{Experiment on Qwen2.5-Coder}\label{subsec:qwen-model-exp}

% \hung{To further validate the findings in this work, we decided to apply Spectral Signature to LLMs with optimized settings based on different attack configurattions. The LLM we choose for this experiment is Qwen2.5-Coder-1.5B, this model has been proven to be SOTA model when it released in coding task, and 1.5B is managable with our current computing resource. On defense side, we used 2 setting, first is the setting defined by the previous work, and the other is the optimized setting based on this work's conducts. The optimized setting, however, were determined based on attack setting including select higher npv and k=50 for 1\% poison rate, in which we assumed that such information is known. On attack side, because of limited resource and based on previous claim to focus on lower poisoning rate \cite{souly2025poisoning}, we will use different attack triggers at 1\% poison rate. The result of this conducted in Table \ref{table:discussion_qwen}, where we report BLEU, ASR-D, and NPV. The setting determined with better result for each attack will be highlighted in italic (lower ASR-D and higher NPV).   As a result, we see that Qwen has minimal effect on Fixed triggers, similar to CodeT5, but the optimized setting has been significantly improved on Grammatical and Adaptive triggers at 50\% and 12.5\% improvement, respectively. Overall, primary result of Qwen2.5-Coder based on previous 3 RQs are relatively correct, the optimized result showed a better result than existing work, and a higher NPV can result a lower ASR-D. }

To further validate the findings in this work, we apply the Spectral Signature defense to a more recent LLM (i.e., Qwen2.5-Coder~\cite{hui2024qwen2}) with optimized settings based on different attack configurations. 
Qwen2.5-Coder~\cite{hui2024qwen2} is a decoder-only language model derived from the Qwen2.5 architecture, specifically designed for coding tasks. It is available in a variety of model sizes, ranging from 0.5B to 32B parameters. Upon its release, Qwen2.5-Coder achieved state-of-the-art performance among open-source code LLMs, matching the coding capabilities of GPT-4o across multiple benchmarks.

The LLM chosen for this additional experiment is Qwen2.5-Coder-1.5B, a model that achieved SOTA performance on coding tasks upon its release, and whose 1.5B parameter size remains manageable with our current computing resources. On the defense side, we employ two settings: the used setting defined by the previous work, and the optimized setting based on the findings from this work. The optimized setting, however, was determined based on the attack configuration, including selecting a higher NPV and $k=50$ for the 1\% poison rate, in which we assumed that such information is known. On the attack side, due to limited resources and based on the previous claim to focus on lower poisoning rates \cite{souly2025poisoning}, we use different attack triggers at the 1\% poison rate. 

The results of this experiment are presented in Table \ref{table:discussion_qwen}, where we report BLEU, ASR-D, and NPV. The setting that yields a better result for each attack is highlighted in italic (lower ASR-D and higher NPV). As a result, we observe that Qwen exhibits minimal effect on \texttt{Fixed} triggers, similar to CodeT5, but the optimized setting shows significant improvement on \texttt{Grammatical} and \texttt{Adaptive} triggers at 50\% and 12.5\% reduction in ASR-D, respectively.

\begin{table}[h]
\centering
\caption{Spectral Signature using Qwen2.5-Coder against 1\% backdoor attack}
\begin{tabular}{cccccc}
\toprule
\textbf{Optimized?} & \textbf{\begin{tabular}[c]{@{}c@{}}Attack\\ method\end{tabular}} & \textbf{k}           & \textbf{ASR-D}   & \textbf{BLEU} & \textbf{NPV}       \\ \midrule
\textbf{No}         & \textit{\textbf{Adaptive}}                                       & \textit{\textbf{10}} & 25.00\%          & 25.01         & 99.9555\%          \\
\textbf{Yes}        & \textit{\textbf{Adaptive}}                                       & \textbf{50}          & \textit{12.50\%} & 26.57         & \textit{99.9557\%} \\
\textbf{No}         & \textit{\textbf{Grammatical}}                                    & \textit{\textbf{10}} & 100.00\%         & 13.20         & \textit{99.9568\%} \\
\textbf{Yes}        & \textit{\textbf{Grammatical}}                                    & \textbf{50}          & \textit{50.00\%} & 12.86         & \textit{99.9568\%} \\
\textbf{No}         & \textit{\textbf{Fixed}}                                          & \textit{\textbf{10}} & 100.00\%         & 13.11         & 99.9076\%          \\
\textbf{Yes}        & \textit{\textbf{Fixed}}                                          & \textbf{50}          & 100.00\%         & 12.89         & \textit{99.9176\%}\\
\bottomrule
\end{tabular}
\label{table:discussion_qwen}
\end{table}

% \subsection{Why CodeBERT is better than CodeT5 for fix and Grammatical trigger while CodeT5 is more suitable for adv?}

% \bowen{come up with some hypothesises}

\subsection{Why SS fails?}

\begin{table}[]
\centering
\caption{Comparison of Characteristics (Code Length, Logical Complexity, and Number of Backdoor Insertions) of True Positive (TP), False Negative (FN), and False Positive (FP) Instances Across Different Poisoning Rates}
\label{tab:why_ss_fail}
\resizebox{\columnwidth}{!}{
\begin{tabular}{@{}ll*{3}{r}@{\hspace{1em}}*{3}{r}@{}}
\toprule
\textbf{Criteria} & \textbf{Rate} & \multicolumn{3}{c}{\textbf{CodeBERT}} & \multicolumn{3}{c}{\textbf{CodeT5}} \\
\cmidrule(lr){3-5} \cmidrule(l){6-8}
 & & \textbf{TP} & \textbf{FN} & \textbf{FP} & \textbf{TP} & \textbf{FN} & \textbf{FP} \\
\midrule

\multirow{3}{*}{\textbf{Length}} 
& 1\% & \textit{92.39} & \textbf{114.21} & 80.13 & \textbf{102.25} & \textit{73.67} & 88.50 \\
& 5\% & 125.70 & \textbf{164.22} & \textit{106.93} & 59.44 & \textit{44.14} & \textbf{72.67} \\
& 10\% & \textit{90.75} & \textbf{118.07} & 99.98 & \textbf{120.11} & \textit{40.60} & 71.77 \\

\midrule

\multirow{3}{*}{\textbf{Complexity}} 
& 1\% & \textit{3.45} & 3.50 & \textbf{3.53} & 2.81 & \textit{2.14} & \textbf{3.18} \\
& 5\% & 3.43 & \textbf{4.61} & \textit{2.96} & \textbf{5.00} & \textit{1.50} & 3.50 \\
& 10\% & \textit{3.18} & \textbf{4.17} & 3.78 & \textbf{4.87} & \textit{1.60} & 2.50 \\

\midrule

\multirow{3}{*}{\textbf{\#Backdoor}} 
& 1\% & \textit{46.75} & \textbf{57.12} & — & \textbf{42.00} & \textit{22.70} & — \\
& 5\% & \textit{49.18} & \textbf{56.56} & — & \textbf{59.87} & \textit{31.20} & — \\
& 10\% & \textit{41.01} & \textbf{44.50} & — & \textbf{43.25} & \textit{26.66} & — \\

\bottomrule
\end{tabular}
}\\
\vspace{1mm}
\textbf{Bolded number} and \textit{italic number} are maximum and minimum number  same code model
% \vspace{-4mm}
\end{table}

% \bowen{extract the poisoning instances that can always bypass the detection.
% 0. identify the overall best configuration of SS, i.e., k=xx, model =yy, poisoning rate = zz, attack = adv.
% 1. collect the specific group of samples.
% 2. manually read a significant portion of samples.
% % use https://www.calculator.net/sample-size-calculator.html?type=1&cl=95&ci=5&pp=50&ps=100&x=Calculate
% 3. come up with hypothesis, what are the unique feature of this set of failing samples.
% 4. verify whether the feature is really \textbf{unique}, we need to measure the feature in the whole dataset as well as the correctly predicted dataset.
% 5. provide and explain a few qualitative samples (optional).
% }

% \bowen{paragraph 1: analysis methodology}

To further understand why SS fails, we sample code snippets misclassified by SS and analyze their characteristics. The code sample snippets are collected from the most advanced (i.e., \texttt{Adaptive} trigger-based) backdoor attack and divided into 3 groups: true positive (i.e., actual poisoned and predicted poisoned), false negative (i.e., actual poisoned but predicted clean), and false positive (i.e., actual clean and predicted poisoned). We focused on the difference between false negative group and the other groups. We hypothesize the correlated characteristic patterns in terms of (1) number of tokens (\textbf{code length}); (2) number of condition clauses and loop clauses (\textbf{code complexity}); (3) number of backdoor attack triggers inserted (\textbf{\#backdoors}). We included both CodeBERT and CodeT5 representation vector for SS and its best configuration against \texttt{Adaptive} trigger in this experiment.

The result of pattern analysis between SS using CodeBERT and CodeT5 in Table \ref{tab:why_ss_fail} has opposite pattern properties from each other. Poison examples that are predicted clean (False Negative) by SS using CodeBERT are, on average, longer code length, more clauses, and have more variables inserted compared to the two other groups (True Positive and False Positive).
% \bowen{what does the two other groups mean here, pls clarify.} \hung{fixed, added TP and FP} 
% \bowen{what do u mean by "number of inserted backdoors" in the false negative group following sent? should not all the false negative group are clean sample but mispredicted as poisoned? } \hung{no that will be false positive}\hung{fixed to trigger}
% On the other hand, the false negative group detected by SS using CodeT5 has a smaller code length, code logic complexity, and number of triggers inserted. 
% \bowen{what does the following sent mean?} \hung{from the table we can see that the representation vector can bias on some visible feature like code length, code complexity, etc}
% \bowen{it's still hard to understand, what do you mean by bias here? can you be more specific or provide some support?}
% \hung{ can u check thiss out, I fixed the writing in next sentence}
% \bowen{it does not make sense because fixed and grammar trigger indeed increase the code length by adding more code, so if we observe clean code with longer length is more likely be predicted as poisoned, then it is reasonable. But here, our results are acutally the opposite. clean code prediced as poisoned because they are shorter. can you explain this?}\hung{thiss include adaptive only}
This result indicates that clean code snippets' representation can be magnified by code properties (such as code length or code complexity), which results in them being mispredicted as poisoned. Therefore, we encourage future researchers to study and develop a specialized representation learning of code that can focus on robustly representing the attack's potential trigger.
% , not influenced by other code's properties.
% \bowen{what is the definition of "good" mean here based on our results?} \hung{i wwould use unbias, my hypothesis is that because the representation vectors seems to bias those properties (longer/shorter code length, more/less complexity) to predict true, it can make the score change base on those feature instead of actual poison trigger.}

\subsection{Findings, Implications and Future Directions}\label{subsec:poison_rate_detection}

% \bowen{@hung, write whatever u think is related and insights drive from our experiments results}
Our experimental results for RQ1 demonstrate that the current use of SS in prior works is often suboptimal. The performance of SS can be significantly improved by identifying a suitable configuration for a given attack scenario. Therefore, it is crucial to guide future research by revealing the patterns of the potential optimal configurations and pointing out the future directions based on our results.

\noindent$\bullet$ \textbf{Is \texttt{Adaptive} trigger always more dangerous than \texttt{Fixed}/\texttt{Grammatical} trigger?}
Yang et al.~\cite{yang2024stealthy} claim that AFRAIDOOR is more stealthy because it is less likely to be detected; however, based on the NPV metric and ASR-D from Section \ref{sec:RQ3_result}, after defense, the poisoned proportion of the remaining dataset is low, and the \texttt{Adaptive} trigger is less effective than \texttt{Fixed} and \texttt{Grammatical} trigger. 
% \bowen{what is the order for the 3 methods here correspond to the later 3 numbers?}\hung{it is the same order}\bowen{I do not really mean to ask u to answer my comment here, instead, i mean pls clarify in the writing to answer my question. The current writing is not really clear about the numbers 99.67, 99.83, and 99.69 correspond to which method. so pls clarify}\hung{is it better?}
% \bowen{what do we mean by "very low poisoning rate" here but we still give 5\% as an example? 5\% is used in \cite{yang2024stealthy}, do you mean with the better configuration of SS instead of using the default one? pls clarify} \hung{ i mean that because 99.67, ... are percentage of clean example of new dataset after clean it, basically they are roughly 0.4\% poison rate}\bowen{pls update the writing to explain what do you mean by "very low"}
% For example, in 3 attack methods at 5\% poisoning rate, the NPV scores of SS at 15\% removal percentage are at 99.67, 99.83, and 99.69, for Adaptive, Fixed, and Grammatical trigger, respectively, while ASR-D score from Table\ref{tab:current_vs_optimal} are 0\%, 28.7\%, and 4.29\% in the same order.
For three attack methods at a 5\% poisoning rate, the NPV scores of the best setting of SS at a 15\% removal percentage are 99.67, 99.83, and 99.69 for \texttt{Adaptive}, \texttt{Fixed}, and \texttt{Grammatical} triggers, respectively, while the ASR-D scores from Table \ref{tab:current_vs_optimal} are 0\%, 28.7\%, and 4.29\%, also in the same order.
% Noted that, NPV also indicates the percentage of clean samples of the dataset after trimmed by SS.
Our results of NPV show that after sample removal, if the remaining dataset has a very low poison percentage (in this case, 0.4\%), the \texttt{Adaptive} trigger is harder to poison the dataset successfully.
% the Adaptive trigger is not without weakness; at a very low poisoning rate such as 0.4\%, it is unlikely to result in a successful attack.

% \hung{fix some accuracy to npv part (include statistic)}
% \hung{Is Adaptive trigger is more dangerous than fix/grammar trigger? Clue: Adaptive trigger need more poison example to be affective although it is harder to detected by ss (accuracy might be needed)}
% \hung{@bowen, i just added this discussion}

\noindent$\bullet$ \textbf{Accurately estimating poisoning rate can make defensive methods more practical and effective, but it currently remains underexplored.} 
While there is no universal configuration for the SS, our results show that the poisoning rate can significantly impact the performance of SS. However, the poisoning rate is often predefined in the existing works, which leads to impractical use of SS as the poisoning rate is unknown from defender's perspective. Thus, it's crucial to develop a method that can accurately estimate the poisoning rate on a given poisoning dataset.
% According to section \ref{sec:ss} performance score and poison percentage have good correlation, we strongly encouraging some further investigation on this relationship. Combine with findings from fig.\ref{graph2.1:asr_CodeBERT_3d} and fig.\ref{graph2.1:asr_CodeT5_3d} that the low poisoning rate group, a high number of eigenvectors in the SS can effectively defend against the attack, vice versa. 
Specifically, distinct configurations may be applied accordingly by leveraging our findings in our results of RQ2 (Section \ref{subsec:RQ2_result}). 
% Based on the result in \fix{Table\ref{RQ2:k_value},\ref{RQ2:models}, and \ref{RQ2:removal rate}} 
%Fig.\ref{graph2.1:asr_CodeBERT_3d} and \ref{graph2.1:asr_CodeT5_3d}
Specifically, for the low poisoning rate group, a high number of eigenvectors in the SS can effectively defend against the attack. On the other hand, high poisoning rate attacks can be defended with a low number of eigenvectors.
Empirically, as shown in Table~\ref{tab1.2:model_performance_poison}, we demonstrate the feasibility of a simple and practical approach to classify attack types based on test-time performance characteristics. In scenarios with high poisoning rates, the performance of a model evaluated on a poisoned test set substantially exceeds the originally reported benchmark or public leaderboards, such as~\cite{bigcodebench_leaderboard,evalplus_leaderboard}.
% Under publicly reported dataset such as CodeSearchNet that has been tested on models like CodeBERT\cite{feng2020codebert}, CodeT5\cite{wang2021codet5}, with high poisoning rates, the performance of a model evaluated on a poisoned test set substantially exceeds the originally reported benchmark.
% \bowen{do you mean "if the evaluation results on the test data is much better than the results on internal test data, then the model developers should be aware the potential of data poisioning?" in the comment 1.9, it seems the COMPARISON is unclear.}\hung{I intend to use the general dataset only, maybe an assumption should been made here. For example, I will assume that there is a dataset known to a group of people and that group also known the standard performance score of some basic models to make it as a reference.}
Thus, \textbf{a substantial performance gain relative to the reported baselines can serve as an indicator of high-rate poisoning}.
In contrast, low poisoning rate attacks typically yield negligible differences between locally evaluated test performance and the originally reported results. Therefore, in the absence of a significant performance discrepancy, it is reasonable to assume a low poisoning rate and to default to a setting optimized for such configurations.

% \hung{On a side note, Souly et al.\cite{souly2025poisoning} also showed that, regardless of the dataset size, only poisoning a small portion of the dataset, sometimes near constant number, can successfully poison models fine-tuned by that dataset. This result can shift the focus of defense setting to detect low percentage poison dataset.}

% \hung{On a side note, Souly et al.~\cite{souly2025poisoning} also demonstrated that, regardless of the dataset size, poisoning only a small portion of the dataset---sometimes a near-constant number of samples---can successfully compromise models fine-tuned on that dataset. This finding further motivates shifting the focus of defense settings toward detecting low-percentage poisoned datasets.}

\noindent$\bullet$ \textbf{NPV: A more reliable proxy metric to approximate the actual defensive performance.} Despite its effectiveness, directly evaluating SS performance through the attack success rate under defense (ASR-D) entails substantial costs for defenders, as it necessitates re-training code models across all configurations. Therefore, it is crucial to identify a reliable and cheaper proxy metric that can approximate the true defensive performance. Prior studies~\cite{ramakrishnan2022backdoors, yang2024stealthy} have commonly employed Recall, the proportion of successfully removed poisoned samples from the dataset, as such a proxy. However, our experimental findings presented in RQ3 reveal that Recall exhibits a weak correlation with actual ASR-D. This observation suggests that recall is an unreliable indicator of SS defense effectiveness and is suboptimal for guiding configuration selection. 
As an alternative, we identify NPV, defined as the proportion of clean samples over predicted clean samples, as a more robust proxy. Our results demonstrate that NPV strongly correlates with ASR-D, making it a preferable metric for approximating defense efficacy. Furthermore, this approximation strategy may extend beyond Spectral Signature to other classification-based defense techniques, given their reliance on outcomes rather than defense-specific features.

% \bowen{\textbf{finding 1} (descrbied in one line), present concrete results, described the corresponding impact if any, how to improve SS}

% \bowen{\textbf{finding 2} (descrbied in one line), present concrete results, described the corresponding impact if any, how to improve SS}

% \bowen{\textbf{finding 3} (descrbied in one line), present concrete results, described the corresponding impact if any, how to improve SS}

% \noindent\textbf{False positives (wrongly predicted poisoning samples, i.e., poisoned predicted as clean)}.

% \noindent\textbf{False negative (wrongly predicted clean samples, i.e., clean predicted as poisoned)}.

% \subsection{batch size impact to SS}

% \bowen{run experiment on different batch sizes}

\subsection{Threats to Validity.} 

\textbf{Threats to Internal Validity} concern potential inaccuracies or errors within the experimental process. One possible threat to our internal validity arises from implementation bugs. To mitigate this risk, we leverage the official repositories of CodeBERT and CodeT5 for both training and inference. For the implementation of SS, we adopted a rigorous development protocol, wherein two authors independently review the code to validate its correctness. Consequently, we believe this threat to be minimal.

\textbf{Threats to External Validity} concern the generalizability of our findings. Previous work~\cite{ramakrishnan2022backdoors} focused solely on outdated models, such as code2vec~\cite{alon2019code2vec} and code2seq~\cite{le2022coderl}, and evaluated only a single type of attack. Differently, our study broadens the scope by incorporating two pretrained models widely used in the code backdoor works~\cite{sun2023backdooring,li2023multi,wan2022you,li2022poison,yang2024stealthy}, namely CodeBERT~\cite{feng2020codebert} and CodeT5~\cite{wang2021codet5}, and examines three recent types of attacks. \fix{Due to resource constraints, we leave the full investigation of larger LLMs for future work.}  Furthermore, our experiments are limited to a single downstream task, code summarization for Python. While this task is widely adopted in prior studies \cite{yang2024stealthy,wan2022you} as a critical attack scenario (e.g., summarize a vulnerable code as safe), the conclusions drawn may not directly transfer to other tasks or programming languages, such as code generation for Java. Differences in downstream tasks and model complexity may influence the effectiveness of Spectral Signature. Future work will therefore extend our evaluation to a broader set of downstream tasks and models to strengthen the generalizability of our findings.
% \fix{C1.15}{Due to resource constraints, we leave the investigation on larger LLMs (such as CodeLlama~\cite{roziere2023code}) as our future work.}
% % However, \fix{C1.15}{because of high computational cost as mention in Section \ref{sec:intro}, some of the more recent LLMs have not been carried out in this paper.} 
% \fix{C1.16}{Furthermore, our experiments are limited to a single downstream task, code summarization. While this task is widely adopted in prior studies \cite{yang2024stealthy,wan2022you} as a critical attack scenario (e.g., summarize a vulnerable code as safe), the conclusions drawn may not directly transfer to other tasks such as code generation and completion. The difference in downstream tasks, model complexity which may influence the effectiveness of Spectral Signature. Future work will therefore extend our evaluation to a broader set of downstream tasks and models to strengthen the generalizability of our findings.}
% Therefore, certain conclusions drawn from this work may not be fully generalizable to other models and attacks. To further reduce this threat, we plan to extend our study in future work by including a wider range of models and datasets.

\textbf{Threats to Construct Validity} concerns the appropriateness of our evaluation metrics. To mitigate this threat, we have carefully selected widely used and well-established evaluation metrics, as outlined in Section~\ref{sec:eval_metrics}. Therefore, we believe the threat is minimal.

% \textbf{Resource limitation.} Lack of resource lead to lack of model diversity (LLMs such as GPT, Llama were not considered), limited configuration can be tested, and limited attack methods can cover.

% \textbf{Correlation to causal.} Although causal often related to high correlation, there are more inspection should be done to verify accuracy and ASR-D relation. For example, other than removal percentage, we should consider more co-founders that can affect the relation such as another defense methods.

\section{Related Works}

\subsection{Code Backdoor Attack}

% \bowen{@justin, pls briefly summarize each attack method}

Existing code attack methods serve specific malicious purposes, such as forcing code models to rank vulnerable code candidates at the top of the list (e.g.,~\cite{sun2023backdooring}), generating misleading code summaries that indicate secure code when it is actually vulnerable (e.g.,~\cite{yang2024stealthy}), or suggesting the use of code that grants remote access to attackers (e.g.,~\cite{schuster2021you}). These methods target a wide range of coding tasks, including code generation~\cite{sun2022coprotector}, code search~\cite{sun2023backdooring,wan2022you,sun2022coprotector}, code translation~\cite{li2023multi}, code summarization~\cite{sun2022coprotector,li2023multi}, code repair~\cite{li2023multi,li2022poison}, code completion~\cite{schuster2021you}, defect detection~\cite{li2022poison,li2023multi}, and clone detection~\cite{li2022poison}. Some methods are designed for specific tasks (e.g.,~\cite{sun2023backdooring,schuster2021you,wan2022you}), while others are task-agnostic (e.g.,~\cite{sun2022coprotector,yang2024stealthy}).
To remain stealthy, existing methods poison code by performing slight perturbations without changing the functional behavior of the original code. For example,~\cite{wan2022you,li2022poison} proposed poisoning the code by adding dead code (as shown in Figure~\ref{fig:examples}.(c) and (d)), such as inserting a \textit{assert} statement like ``\textit{$assertTrue(1\geq0);$}'' or inserting a variable declaration like ``\textit{$int\ ret\_Val\_;$}''. Code renaming is also a common strategy used in~\cite{sun2023backdooring,sun2022coprotector,yang2024stealthy} (as shown in Figure~\ref{fig:examples}.(b)). More specifically, existing methods propose to rename different identifiers in the code, such as method names~\cite{li2022poison}, variable names~\cite{li2022poison,sun2022coprotector}, API names~\cite{sun2022coprotector}, or a mixture~\cite{sun2023backdooring,yang2024stealthy}.
% However, there are some exceptions.~\cite{li2023multi} proposed to poison the dataset by adding code files without changing the original code.~\cite{sun2022coprotector} proposed to replace the statements in the original code with the same type of statements randomly chosen from other functions in the same repository.
% (2) \textbf{Trigger generation strategy.} As shown in Figure~\ref{fig:examples}, 

% \begin{figure}{r}{4in}
% \includegraphics[width=\columnwidth]{imgs/examples.pdf}
%  \caption{Examples of the Fixed~\cite{li2022poison,wan2022you}, \textit{grammatical}~\cite{sun2022coprotector,wan2022you,li2023multi}, Adaptive triggers~\cite{yang2024stealthy}. Changes to the original function are highlighted in \colorbox{yellow}{yellow}.}
%  \label{fig:trigger}
%  % \vspace{-1em}
% \end{figure}

\subsection{Other Defense Methods}

Besides Spectral Signature (SS), we present additional backdoor defense techniques that have been explored in the literature: 
\fix{"ONION  \cite{qi2021onion} defends against backdoor attacks by identifying outlier tokens that most affect sentence perplexity; in code, it targets trigger variables. Activation Clustering \cite{chen2018detecting} applies PCA to the last hidden layer and clusters representations to separate clean and poisoned data."} \hung{trimmed here}
% \textbf{ONION} was proposed by Qi et al. \cite{qi2021onion} to defend against backdoor attacks on text. The original work idea is to find outliners that affect the perplexity $p_0$ of a sentence $d=w_1,w_2, ..., w_n$ the most. The score of a word $w_i$ is computed by the difference between $p_0$ and $p_i$ where $p_i$ is the perplexity of the sentence $d\setminus w_i$. In the code context, ONION aims to remove variables that are used to trigger the output of the attack method. \textbf{Activation clustering} was proposed by Chen et al. \cite{chen2018detecting} to use the output of the last hidden layer of the code model, apply Principal Component Analysis (PCA)\cite{abdi2010principal} to it, and use a clustering method to divide clean and poisoned data.
In this study, we exclusively focus on the Spectral Signature (SS) approach due to its widespread adoption and demonstrated effectiveness in prior work, which demonstrates its superior performance compared to other existing backdoor defense methods~\cite{ramakrishnan2022backdoors}.

\section{Conclusion}

In this paper, we re-evaluate and study the Spectral Signature comprehensively in various configurations against state-of-the-art code models and backdoor attacks. Our empirical evaluation unveils that the previously used configuration of SS is not optimal in over 66.67\% of attack scenarios. Moreover, we found that the optimal configuration of SS for the number of top \textit{k} eigenvectors is related to the poisoning rate, where high \textit{k} ($\geq10$) suitably defends against lower poison rates of the attack, and low \textit{k} ($\leq3$) is more effective in mitigating higher poison rate attacks. Finally, we compared the Recall, FPR, and NPV evaluation metrics in relation to ASR-D using Spearman and Kendall $\tau$ correlation. The experimental results show that the NPVs' correlation to the practical evaluation of SS is consistently higher than recalls in both measurements, 
% (0.377 and 0.302  average difference in Spearman and Kendall $\tau$)
suggesting that future work should adopt NPV as the primary proxy metric.
In the future, we plan to expand the study by experimenting with more code models and attack methods.

\section*{Data Availability}

To promote transparency and facilitate reproducibility, we make our artifacts available to the community at:     \url{https://github.com/hunglequoc2001/poisoning-attack-for-LLMs}
This repository includes a replication package of Spectral Signature and datasets used in our experiments, and the scripts for results analysis.

\bibliographystyle{IEEEtran}
\bibliography{main}

% \begin{thebibliography}{99}
% % paste content of main.bbl here
% \end{thebibliography}

\end{document}